\documentclass{aa}

\usepackage{amsmath}
\usepackage{bm}
\usepackage{gensymb}
\usepackage{graphicx}
\usepackage{natbib}
\usepackage{placeins}
\usepackage{subcaption}
\usepackage[flushleft]{threeparttable}
\usepackage{txfonts}
\usepackage{textcomp}
\usepackage{upgreek}
\bibpunct{(}{)}{;}{a}{}{,}

\begin{document}

   \defcitealias{molina18}{MB18}

   \title{A model for high-mass microquasar jets under the influence of a strong stellar wind}
   \titlerunning{HMMQ jets under the influence of a strong stellar wind}

   \author{E. Molina \inst{1} \and S. del Palacio \inst{2} \and V. Bosch-Ramon \inst{1}}

   \institute{Departament de F\'isica Qu\`antica i Astrof\'isica, Institut de Ci\`encies del Cosmos (ICCUB),
              Universitat de Barcelona (IEEC-UB), Mart\'i i Franqu\`es 1, 08028 Barcelona, Spain \\
              \email{emolina@fqa.ub.edu, vbosch@fqa.ub.edu}
              \and Instituto Argentino de Radioastronom\'ia (CCT-La Plata, CONICET; CICPBA),
              C.C.5, 1894 Villa Elisa, Argentina \\
              \email{sdelpalacio@iar.unlp.edu.ar}}

   \date{Received -; accepted -}

  \abstract 
   {High-mass microquasars (HMMQs) are systems from which relativistic jets are launched. At the scales of several times the binary system size, the jets are expected to follow a helical path caused by the interaction with a strong stellar wind and orbital motion. Such a trajectory has its influence on the non-thermal emission of the jets, which also depends strongly on the observing angle due to Doppler boosting effects.}
   {We explore how the expected non-thermal emission of HMMQ jets at small scales is affected by the impact of the stellar wind and the orbital motion on the jet propagation.}
   {We studied the broadband non-thermal emission, from radio to gamma rays, produced in HMMQ jets up to a distance of several orbital separations, taking into account a realistic jet trajectory, different model parameters, and orbital modulation. The jet trajectory is computed by considering momentum transfer with the stellar wind. Electrons are injected at the position where a recollimation shock in the jets is expected due to the wind impact. Their distribution along the jet path is obtained assuming local acceleration at the recollimation shock, and cooling via adiabatic, synchrotron, and inverse Compton processes. The synchrotron and inverse Compton emission is calculated taking into account synchrotron self-absorption within the jet, free-free absorption with the stellar wind, and absorption by stellar photons via pair production.}
   {The spectrum is totally dominated by the jet over the counter-jet due to Doppler boosting. Broadband emission from microwaves to gamma rays is predicted, with radio emission being totally absorbed. This emission is rather concentrated in the regions close to the binary system and features strong orbital modulation at high energies. Asymmetric light curves are obtained owing to the helical trajectory of the jets.}
   {The presence of helical shaped jets could be inferred from asymmetries in the light curves, which become noticeable only for large jet Lorentz factors and low magnetic fields. Model parameters could be constrained if accurate phase-resolved light curves from GeV to TeV energies were available. The predictions for the synchrotron and the inverse Compton radiation are quite sensitive of the parameters determining the wind-jet interaction structure.}

   \maketitle

\section{Introduction}

Microquasars are binary systems consisting of a compact object (CO), either a black hole or a neutron star, that launches relativistic jets powered by accretion of matter from a companion star. When the companion is a massive star, these systems are called high-mass microquasars (HMMQs) and accretion onto the CO takes place as the latter captures a fraction of the stellar wind. This mechanism is compatible with the presence of an accretion disk around the CO \citep{elmellah19a}, which is a necessary condition for jet formation. In HMMQs, interaction between the jets and the stellar wind may play an important role in both the propagation and the radiation produced by the jet, as the combined effect of the wind and orbital motion deviates the jets from a straight trajectory. Several works study the dynamical influence of the stellar wind on the jets of HMMQs, and their expected radiative output, at the scales of the binary system \citep[e.g.,][]{romero05,khangulyan08,perucho08,araudo09,owocki09,perucho10,perucho12,yoon16,khangulyan18}.

At larger scales, the effect of orbital motion becomes important and could make the jets follow a helical trajectory \citep[see, e.g.,][for a semi-analytical study about this]{bosch16}. Current observations allow for a detection of this kind of pattern in microquasars. The most famous case is perhaps SS433, in which the helical structure is likely caused by precession of the accretion disk \citep{begelman06,monceau2014}. Jet precession has also been observed in 1E~1740.7$-$2942 \citep{luque15}, even though it is a low-mass microquasar in which the stellar wind is not expected to have a strong dynamical influence. Cygnus~X-3 is another system for which helix-like jets have been found \citep{mioduszewski01,miller04}, although the cause of their shape remains unclear. Finally, one may expect a strong wind-jet interaction in Cygnus~X-1, given that it hosts a massive star close to the CO \citep[e.g.,][]{yoon16,bosch16}. Nevertheless, observations have not shown evidence of the presence of helical jets in Cygnus~X-1 so far \citep{stirling01}.

Since the orbital motion could leave a strong imprint on jet radiation in HMMQs, \citet[][hereafter MB18]{molina18} computed the non-thermal radiative output of helical jets along the orbit using a phenomenological prescription for the jet kinematics. That work focused on a jet region rather far from the binary system, and the jet speed was taken well below the speed of light, due to a braking effect caused by instability growth and subsequent mixing of jet and wind material \citep{bosch16}. However, closer to the launching site, within a few orbital separations from the jet base, the jet is thought to be relativistic \citep[e.g.,][]{fender04}. In that case, absorption processes and radiation cooling become more important, and Doppler boosting should be considered to compute the observable radiation.

The work presented here complements what was done in \citetalias{molina18} by studying in detail the wind-jet interaction on the scales of the binary system and its peripheral region. For that purpose, the jet trajectory is computed accounting for orbital motion and the momentum transferred by interaction with the stellar wind. The leptonic jet radiation is computed using a semi-analytical code, modeling the jets as one-dimensional emitters that radiate via the synchrotron and inverse Compton (IC) mechanisms. Absorption in radio and gamma rays is also accounted for.

The paper is organized as follows: In Sect.~\ref{scenario}, we describe the physical system under study, as well as the the technical aspects of the dynamical and radiative models adopted. In Sect.~\ref{results}, we present the main results of this work. Finally, a summary and a discussion of the results are given in Sect.~\ref{discussion}.
 
\section{Model description}\label{scenario}

\subsection{System properties}\label{system}

\begin{figure}
    \centering
    \includegraphics[width=0.95\linewidth]{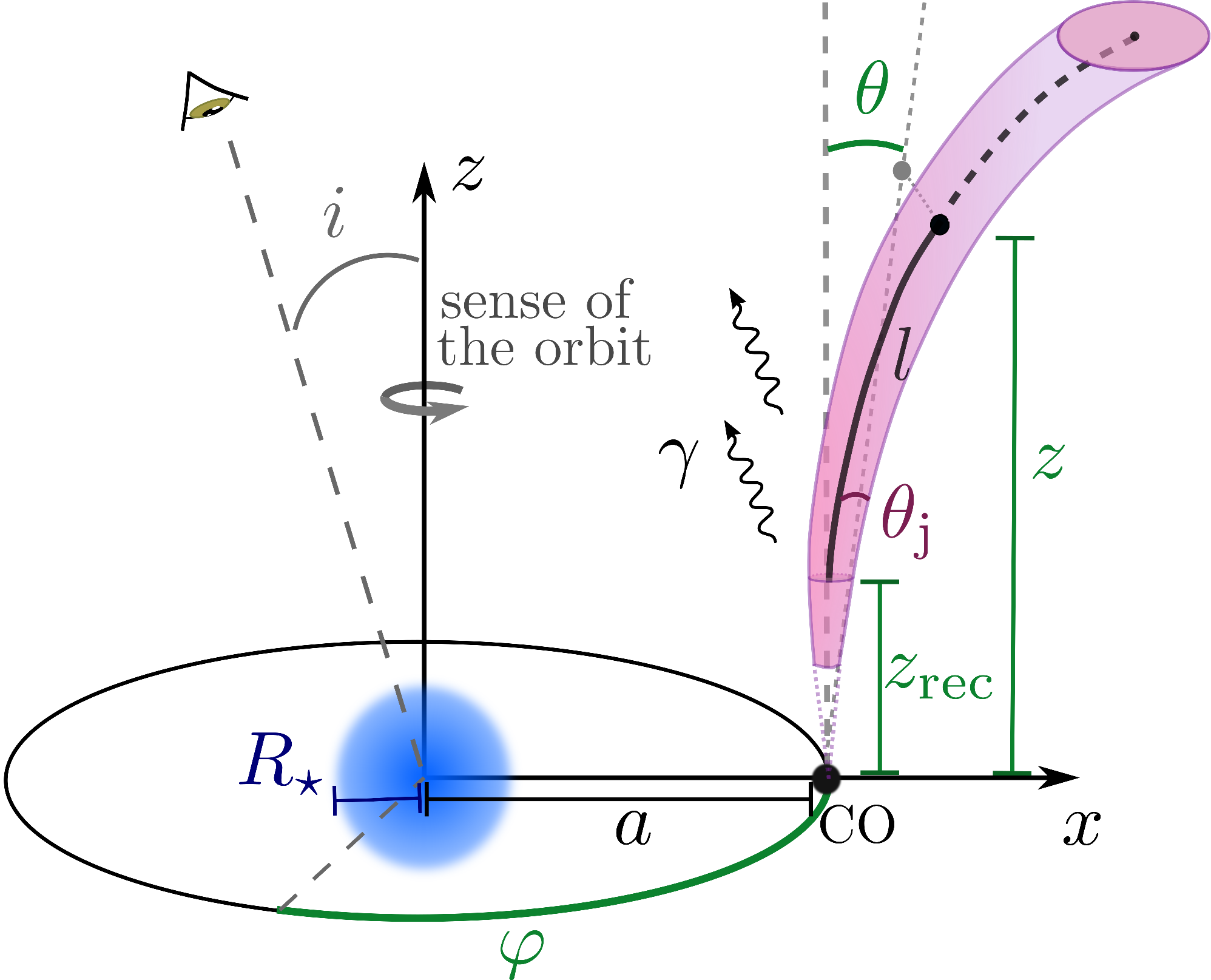}
    \caption{Sketch of considered scenario (not to scale). We show a jet segment at a height $z$ (black dot) and its projection in the $xz$-plane (grey dot). We also show the recollimation height $z_{\rm rec}$, the length $l$ along the jet, the jet half-opening angle $\theta_\mathrm{j}$, the jet bending angle with respect to the $z$-axis ($\theta$), the inclination angle of the orbit $i$, the orbital separation $a$, the star radius $R_\star$, and the orbital phase $\varphi$ (set to $0$ and $0.5$ for the inferior and superior conjunctions, respectively). The counter-jet (not shown) is symmetric to the jet with respect to the $xy$-plane.}
    \label{fig:sketch}
\end{figure}

The system studied in this work is a generic HMMQ, similar to Cygnus~X-1, in which a CO is accreting matter from a massive companion star. The orbit is considered to be circular, with a period of $T = 4$~days and a separation of $a = 3\times10^{12}$~cm $\approx 0.2$~AU. A jet and a counter-jet are ejected in opposite directions perpendicularly to the orbital plane, and progressively deviate from their initial direction by the combined effects of the stellar wind and orbital motion. The star is located at the origin of our coordinate system, which co-rotates with the CO counter-clockwise. For simplicity, we assumed that the star rotates synchronously with the coordinate system. The initial direction of the jets is taken as the $z$-axis. The $x$-axis is defined by the star-CO direction and the $y$-axis is perpendicular to it, in the direction of the orbital motion. The orbital phase of the CO is characterized by $\varphi$. The system is assumed to be at a distance of $d = 3$~kpc from an observer that sees the system with an inclination $i$ with respect to the $z$-axis. Figure~\ref{fig:sketch} illustrates this scenario.

The companion is a typical O-type star in the main sequence, with a temperature of $T_\star = 40\,000$~K and a luminosity of $L_\star = 10^{39}$~erg~s$^{-1}$ \citep{heap06}, which correspond to a radius of $R_\star \approx 7.4\times10^{11}$~cm. The stellar wind is assumed isotropic, with a velocity following the usual \textit{$\beta$-law} form $v_{\rm w} = v_\infty(1-R_\star/r)^{\hat\beta}$, where $v_\infty = 2\times10^8$~cm~s$^{-1}$ is the terminal wind speed, $r$ is the distance to the star, and $\hat\beta = 0.8$ is typical for hot massive stars \citep[e.g.,][]{pauldrach86}. Although the $\hat\beta$-value is not tightly constrained, our results are almost insensitive to it within its usual range for O-type stars, $0.7 \lesssim \hat\beta \lesssim 1$ \citep{puls96}, and not very sensitive to values of $\hat\beta$ up to 3 found for some stars with different spectral types \citep{crowther06}. However, if a smaller $a$ was considered, a variation of the $\hat\beta$-value would have a higher impact in our results given the proximity of the jets to the star. We also note that the adopted $v_\infty$ is typical for O-type stars, but can be significantly lower for those with a later spectral type ($\sim$ 500 km s$^{-1}$ for the B-supergiant in Vela X-1; \citealt{sander18}). The combination of a larger $\hat\beta$ and a smaller $v_\infty$ for stars with a later spectral type would likely make their wind influence on the jets negligible. The mass-loss rate of the stellar wind is taken as $\dot{M}_{\rm w} = 10^{-6}$~M$_\sun$~yr$^{-1}$, which is also a characteristic value for O-type stars \citep{lamers99,muijres12}.

As shown in Fig.~4 of \cite{yoon16}, for a case similar to that studied here, the geometry of a jet shocked by the stellar wind departs from that of a conical jet. On the other hand, the shocked material surrounding the jet also takes part in the wind-jet momentum transfer, although this effect is difficult to capture without detailed simulations. Thus, for simplicity, we assume that the jets have a conical shape of half-opening angle $\theta_{\rm j} = 0.1$~rad, since they have been shown to be significantly collimated \citep[e.g.,][]{stirling01}. The jets are assumed to have a constant Lorentz factor $\gamma_{\rm j}$ from their launching to end points, the latter located at several $a$. Two values of $\gamma_{\rm j}$ are considered in order to assess the impact of this parameter in the results (e.g., due to different Doppler boosting).

Both jet and counter-jet have a kinetic power of $L_{\rm j} = 5\times10^{36}$~erg~s$^{-1}$, a fraction $\eta_{\rm NT}$ of which is injected into non-thermal electrons accelerated with a rate $\dot{E}_{\rm acc}=\eta_{\rm acc} e c B$, where $\eta_{\rm acc}$ is the acceleration efficiency, and $e$ the elementary charge. The fiducial values of $\eta_{\rm NT}$ and $\eta_{\rm acc}$ are both set to 0.1, but they are not well constrained beyond the fact that they both should be $\le 1$ (see Sect.~\ref{discussion}). We focus here on electrons (and positrons) as hadronic radiation processes are far less efficient than leptonic ones under the conditions assumed in the emitting regions \citep[e.g.,][]{bosch09a}. We characterize the magnetic field at the jet base through a fraction $\eta_B$ of the total jet energy density (see Sect.~\ref{partcool}). Table~\ref{tab:parameters} summarizes the different parameters that are used in this work.

\begin{table}
 \begin{center}
	\caption{List of the jet, star and system parameters that are used throughout this work. The last three are free parameters for which different values are explored.}
	\begin{tabular}{l c c}
    \hline \hline
	\multicolumn{1}{c}{Parameter}	&			    	&	Value					    	        \\
    \hline
	Stellar temperature				& $T_\star$			&	$4\times10^4$ K 				        \\
    Stellar luminosity 				& $L_\star$ 		& 	$10^{39}$ 	erg~s$^{-1}$		        \\
    Mass-loss rate 			        & $\dot{M}_{\rm w}$ &	$10^{-6}$ M$_\sun$~yr$^{-1}$	        \\
    Terminal wind speed 		    & $v_\infty$ 		&	$2\times10^8$~cm~s$^{-1}$		        \\
    \textit{$\beta$-law} exponent   & $\hat\beta$       &   $0.8$                                   \\
    Jet luminosity 					& $L_{\rm j}$ 		&	$5\times10^{36}$ erg~s$^{-1}$	        \\
    Non-thermal energy fraction 	& $\eta_{\rm NT}$ 	&	$0.1$						            \\
    Acceleration efficiency         & $\eta_{\rm acc}$  &   $0.1$                                   \\
    Jet half-opening angle 			& $\theta_{\rm j}$ 	&	$0.1$~rad						        \\
    Orbital separation 				& $a$ 		        &	$3\times10^{12}$ cm 			        \\
    Orbital period 					& $T$ 				&	$4$~days						        \\
    Distance to the observer 		& $d$ 				&	$3$~kpc 						        \\
    Jet Lorentz factor              & $\gamma_{\rm j}$  &   $1.2$ , $3$                             \\
    Magnetic pressure fraction      & $\eta_B$          &   $10^{-4}$ , $10^{-2}$ , $1$             \\
    System inclination              & $i$               &   $0\degree$ , $30\degree$ , $60\degree$  \\
    \hline
    \end{tabular}
    \label{tab:parameters}
 \end{center}
\end{table}

\subsection{Jet dynamics}\label{hydrodynamics}

\begin{figure}
    \centering
    \includegraphics[angle=270, width=\linewidth]{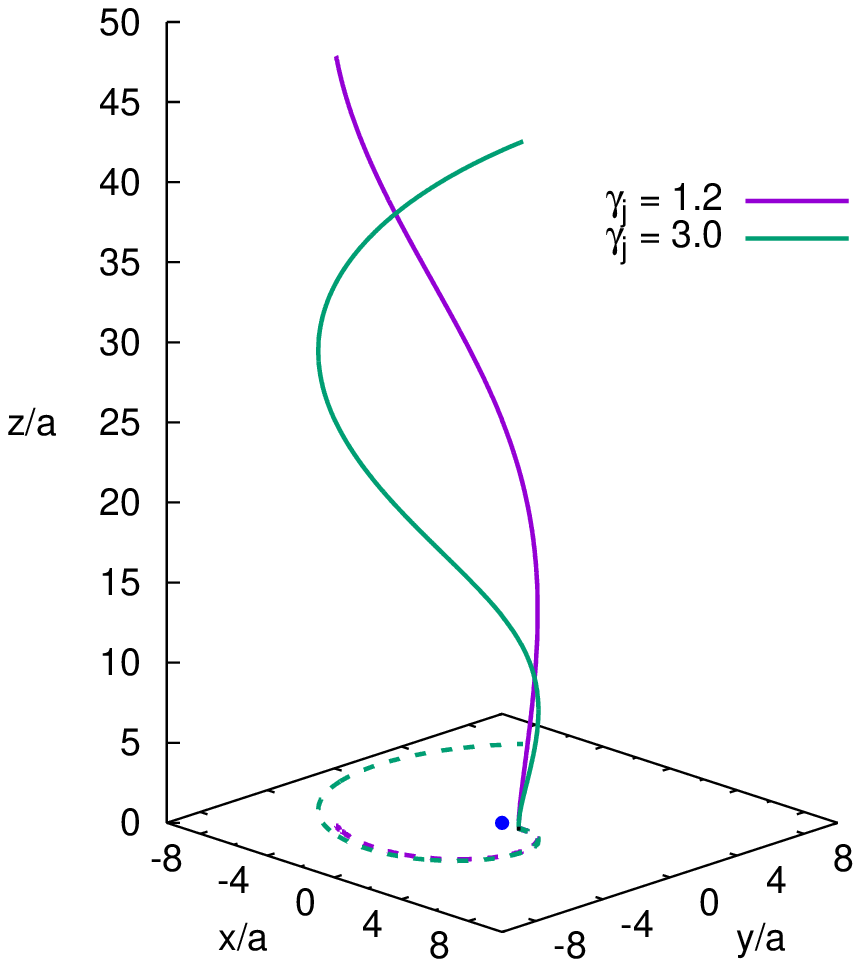}
    \caption{Trajectory followed by the jet for $\gamma_{\rm j} = 1.2$ (solid purple line) and $3$ (solid green line). The projections in the orbital plane are also shown with dashed lines. The jet length is 50$a$, and the counter-jet is symmetric to it with respect to the $xy$-plane (not shown). The blue circle at the origin of coordinates represents the star (to scale). The jets are plotted starting from the corresponding recollimation shocks.}
    \label{fig:trajectory}
\end{figure}

The trajectory of the jet is computed starting from a height $z_0 = 2\times10^{10}$~cm, small enough so that wind effects are negligible. The counter-jet starts at $-z_0$. Initially the jets propagate in the $\hat{z}$ direction. However, the interaction with the stellar wind bends them away from the star in the $\hat{x}$ direction within the scales of the binary system. Additionally, at larger scales, the Coriolis force related to the orbital motion makes the jets bend in the $-\hat{y}$ direction, opposite to the sense of the orbit. We obtain the trajectory of the jets by dividing them into segments and computing iteratively how they are reoriented due to the momentum transfer by the stellar wind. In cylindrical coordinates, the first jet segment sets the following initial conditions for the position $\vec{r}$ and the momentum $\vec{P}$:
     \begin{equation}\label{ICond}
     \begin{aligned}
     \vec{r_1} &= (r,\phi,z) = (a,\varphi,z_0) \ , \\
     \vec{P_1} &= (P_r,P_\phi,P_z) = (0,0,\dot{P_{\rm j}} {\rm d}t) \ ,
     \end{aligned}
     \end{equation}
where $\dot{P}_{\rm j} = L_{\rm j} \gamma_{\rm j} \beta_{\rm j}/c(\gamma_{\rm j}-1)$ is the total jet thrust, $\beta_{\rm j} = v_{\rm j}/c$ is the jet propagation velocity in units of the speed of light $c$, and d$t$ is the segment advection time. In order to get the initial conditions for the counter-jet trajectory one just needs to change the sign of the $z$ coordinate in both $\vec{r_1}$ and $\vec{P_1}$. The Coriolis plus wind forces acting on each segment are:
     \begin{equation}\label{Fw}
     \begin{aligned}
     F_r &= S_\star \rho_{\rm w} v_{\rm w}^2 \cos{\theta} \ , \\
     F_\phi &= \rho_{\rm w} S_\phi \min \left(\frac{4\pi (r-a)}{T} , \frac{2\pi r}{T} \right)^2 \ , \\
     F_z &= S_\star \rho_{\rm w} v_{\rm w}^2 \sin{\theta} \ ,
     \end{aligned}
     \end{equation}
where $\rho_{\rm w}$ is the wind density, $\theta$ is the angle between the $r$ and $z$ coordinates of the segment, and $S_\star$ and $S_\phi$ are the segment surfaces perpendicular to the \boldsymbol{$x$} and \boldsymbol{$\phi$} directions, respectively. As we work under the assumption of an isotropic, spherically symmetric wind, $\rho_{\rm w} = \dot{M}_{\rm w} / 4\pi v_{\rm w} d_\star^2$, with $d_\star = \| \vec{r} \|$ being the distance to the star. Nevertheless, it is worth mentioning that some level of wind beaming towards the accretor is expected, especially for slow winds \citep{friend82,gies86,elmellah19b}. This would increase the wind density in the orbital plane with respect to the isotropic case, while decreasing it off the plane, thus reducing the wind influence on the jets farther from the CO. Nonetheless, we neglect wind beaming since we work with a simplified prescription and a fast stellar wind is considered. The first term in the $\min$ function is the velocity corresponding to the Coriolis force at each segment position, while the second term is the wind velocity in the $\phi$ direction as seen from the jet, as the wind-jet relative $\phi$-velocity cannot be larger than the wind tangential velocity in the non-inertial frame\footnote{We recall that we are assuming a star rotating synchronously with the CO, and thus the wind tangential velocity associated with the stellar rotation is zero.}. The forces in Eq.~\eqref{Fw} are then used to compute the momentum of the subsequent segments as:
     \begin{equation}\label{Pi}
     \vec{P_{\rm i+1}} = \vec{P_{\rm i}} + \vec{F_{\rm i}} \ {\rm d}t \ .
     \end{equation}
As we are taking here a constant propagation velocity $v_{\rm j}$, the additional momentum that this prescription generates is assumed to go into heat of the shocked structure made of interacting jet and wind material. However, we do not consider the back reaction of the accumulated heat, nor instability growth and mixing, on the interacting flows. A more accurate and realistic account of the process would require carrying out costly 3-dimensional, relativistic hydrodynamical simulations. For simplicity, we use a phenomenological approach to compute the jet trajectory, acknowledging that as the jet segments get farther from the binary, our model becomes less realistic. Nevertheless, as the most relevant emitting regions are close to the binary system (see Sect.~\ref{results}), we consider our approach a reasonable one at this stage. For illustrative purposes, Fig.~\ref{fig:trajectory} shows the path followed by the jet for $\gamma_{\rm j} = 1.2$ and $3$.

\subsection{Particle cooling}\label{partcool}

In this work, primed quantities refer to quantities in the fluid frame (FF), whereas unprimed quantities refer to the laboratory frame (LF). The jet emission is computed as in \citetalias{molina18}, but accounting for relativistic effects and additional absorption mechanisms. The electron distribution and the synchrotron and IC luminosities are calculated first in the FF at each point along the jets. Then, the luminosities are computed as seen by the observer taking into account Doppler (de)boosting. Finally, these luminosities are corrected by different absorption processes.

The point where the non-thermal electrons are injected is located at a height $z_{\rm rec}$ ($-z_{\rm rec}$ for the counter-jet) for which a recollimation shock is formed due to the jet and wind momentum fluxes balancing each other in the $y$-axis direction (perpendicular to the jet propagation). This condition can be approximately expressed as follows \citep{bosch16}:
    \begin{equation}\label{recollimation}
    \frac {\dot{P}_{\rm j}}{\pi z_{\rm rec}^2 \gamma_{\rm j}^2} = \frac {\dot{P}_{\rm w}}{4\pi a^2}
    \left( \frac {a^2}{a^2 + z_{\rm rec}^2} \right ) \ ,
    \end{equation}
where $\dot{P}_{\rm w} = \dot{M}_{\rm w} v_{\rm w}$ is the wind thrust. For $\gamma_{\rm j} = 1.2$ we obtain $z_{\rm rec} = 1.4\times 10^{12}$~cm $\approx 0.47 a$, whereas for $\gamma_{\rm j} = 3$, $z_{\rm rec} = 3.1\times 10^{11}$~cm $\approx 0.10 a$. Some combinations of $\dot{P}_{\rm w}$, $\dot{P}_{\rm j}$ and $\gamma_{\rm j}$ yield non-physical (complex) values of $z_{\rm rec}$ in Eq.~\eqref{recollimation}, which means that for those sets of parameters no recollimation shock is formed, as the momentum flux balance is never reached \citep[see][for a non-relativistic simulation where this effect is studied]{yoon16}.

Electrons are injected at $\overrightarrow{r_0} \approx (a,0,\pm z_{\rm rec})$, since the jet displacement in the $x$ and $y$ coordinates is small at $z = \pm z_{\rm rec}$ ($\lesssim 0.005 a$ in the worst case). The particles are assumed to follow a power-law energy distribution with spectral index $-2$, typical for acceleration in strong shocks via the Fermi I mechanism \citep[e.g][]{drury83}, with a minimum energy of $E'_{\rm min} = 1$~MeV. An exponential cutoff is also assumed:
    \begin{equation}\label{Q}
    Q'(E') \propto E'^{-2} \exp{\left( -\frac{E'}{E'_{\rm cutoff}}\right)} \ ,
    \end{equation}
where $E'$ is the electron energy in the FF. The normalization of $Q'(E')$ is obtained from the condition that the total power available for non-thermal electrons in the FF is $L'_{\rm NT} = \eta_{\rm NT} L_{\rm j}/\gamma_{\rm j}^2$. Electrons are convected from $\overrightarrow{r_0}$ along the jet path while they cool down via adiabatic, synchrotron and IC processes. For simplicity, we neglect particle acceleration beyond $\overrightarrow{r_0}$, even though weak shocks and turbulence may (re)accelerate particles further downstream in the jets. Analytical expressions are used for the energy losses in the FF (see \citealt{longair81} and \citealt{khangulyan14} for synchrotron and IC losses, respectively). The characteristic timescale for adiabatic cooling is $t'_{\rm ad} = E'/\dot{E}'_{\rm ad} = 3 R_{\rm j}/2 \theta_{\rm j} v_{\rm j} \gamma_{\rm j}$, where $R_{\rm j}(l)\propto l$ is the jet radius.

The magnetic pressure at the accelerator (located at $\overrightarrow{r_0}$) is characterized in the FF as a fraction $\eta_B$ of the total jet energy density:
    \begin{equation}\label{etaB}
    \frac{B_0^{\prime 2}}{8\pi} = \eta_B \frac{L_{\rm j}}{\pi \gamma_{\rm j}^2 r_{\rm j,0}^2 v_{\rm j}} \ ,
    \end{equation}
where $B'_0 = B_0/\gamma_{\rm j}$ and $r_{\rm j,0} = R_{\rm j}(z_{\rm rec})$ are the magnetic field and the jet radius at $\overrightarrow{r_0}$, respectively. For the same $B'_0$, this fraction relates to the ratio of magnetic pressure to stellar photon energy density in the LF as $\eta_{B}^\star = 4 (r_0/r_{\rm j,0})^2 (L_{\rm j}/L_\star) \beta_{\rm j}^{-1} \eta_B$, which is useful to compare the expected radiation outputs of synchrotron and IC processes. This yields $\eta_B^\star \approx 19\eta_B$ for $\gamma_{\rm j} = 1.2$, and $\eta_B^\star \approx 200\eta_B$ for $\gamma_{\rm j} = 3$. The jet magnetic field is assumed perpendicular to the flow in the FF, since at the scales considered in this work, far from the jet launching point, a dominant toroidal component for the magnetic field is expected \citep[e.g.,][]{pudritz12}. Given the adopted assumptions plus frozen in conditions, the magnetic field at each jet point can be computed as:
    \begin{equation}\label{BFF}
    B'(l) = B'_0 \frac{r_{\rm j,0}}{R_{\rm j}(l)} \ .
    \end{equation}

The cutoff energy $E'_{\rm cutoff}$ is the maximum energy that the electrons can achieve in the accelerator region in the FF, and it is obtained as follows:

\begin{enumerate}
    \item We compare the acceleration timescale $t'_{\rm acc} = E'/\dot{E}'_{\rm acc} = E'/\eta_{\rm acc} e c B'_0$ with the cooling timescale $t'_{\rm loss} = E'/\dot{E'}$, where $\dot{E'}$ is the cooling rate accounting for adiabatic, synchrotron and IC losses. In this work we set $\eta_{\rm acc} = 0.1$ as a representative case with efficient particle acceleration. The energy $E_{\rm max}^{\prime{\rm acc}}$ at which the two timescales are equal is the maximum energy that electrons can attain before the energy losses overcome the energy gain by acceleration.
    
    \item We compare $t'_{\rm acc}$ with the diffusion timescale $t'_{\rm diff} = r_{\rm j,0}^2 / 2 D$, where $D = E' c / 3 e B'_0$ is the diffusion coefficient in the Bohm regime. If diffusion is a dominant process, particles cannot reach $E_{\rm max}^{\prime{\rm acc}}$ before escaping the accelerating region. We can estimate the maximum energy attainable by the electrons before they diffuse away from the accelerator, $E_{\rm max}^{\prime{\rm diff}}$, by equating $t'_{\rm acc} = t'_{\rm diff}$.
    
    \item We obtain the cutoff energy as $E'_{\rm cutoff} = \min(E_{\rm max}^{\prime{\rm acc}} , E_{\rm max}^{\prime{\rm diff}})$. For the different sets of parameters considered in this work we always obtain a $E'_{\rm cutoff}$ of a few TeV.
\end{enumerate}

In order to compute the electron energy distribution $N'(E',l)$ along the jet, the latter is divided into 1500 segments with size $\mathrm{d}l = 10^{11}$~cm, extending to a total length of $l_{\rm j} = 1.5\times10^{14}$~cm $= 50a$. Once the injection function $Q'(E')$ is determined, the electron energy distribution $N'(E',l)$ is computed for each segment $l$ in an iterative way as done in \citetalias{molina18}, but considering that all quantities must be expressed in the FF. This defines a structured, linear relativistic emitter along the trajectory shown in Fig.~\ref{fig:trajectory}.

\subsection{Radiation}\label{radiation}

We focus our study in the computation of synchrotron and IC radiation of leptons along the jets. Hadronic processes are likely much less efficient at the scales considered in this work, given the conditions of the emitting regions. Relativistic bremsstrahlung is also expected to be unimportant in comparison to the synchrotron and IC mechanisms \citep{bosch09a}.

At each segment of the jet, the spectral energy distribution (SED) of the emitted synchrotron radiation for an isotropic population of electrons in the FF is given by \citep{pacholczyk70}:
    \begin{equation}\label{Lsyn}
    \varepsilon' L_{\varepsilon'}^{\prime{\rm syn}}(l) = \varepsilon' \frac{\sqrt{2} e B'(l)}
    {m_{\rm e} c^2 h} \int_{E_{\rm min}}^\infty F(x') N'(E',l) \ {\rm d}E' \ ,
    \end{equation}
where $\varepsilon'$ is the photon energy, $m_{\rm e}$ is the electron mass, $h$ is the Planck constant, $x' = \varepsilon'/\varepsilon'_{\rm c}$, and $\varepsilon'_{\rm c} = \sqrt{3} e h B' E^{\prime 2}/(2 \! \sqrt{2} \pi m_{\rm e}^3 c^5)$ is the critical photon energy assuming an isotropic distribution of the pitch angle $\theta'_B$ between the electron velocity and the magnetic field, such that $B'\!\sin{\theta'_B}\!\approx\!\sqrt{\langle B^{\prime 2} \rangle} = B'\!\sqrt{2/3}$. The function $F(x')$ can be approximated by $F(x') \approx 1.85\ x^{\prime -1/3} e^{x'}$ for $x' \in [0.1,10]$ \citep{aharonian04}.

For the calculation of the IC emission we only consider the up-scattering of stellar photons; other radiation fields (such as the one from the accretion disk) are negligible in comparison to the stellar one on the relevant scales. Synchrotron self-Compton also turns out to be negligible. We use the prescription developed by \cite{khangulyan14} to compute the interaction rate of an electron with a monodirectional field of target photons with a black body distribution, ${\rm d}^2 N'/{\rm d}\varepsilon'{\rm d}t'$. For simplicity, we do not consider possible deviations from a black-body spectrum owing to absorption of stellar photons by wind material, since this is not expected to have a significant effect on the IC emission \citep[see][for an illustrative comparison between a black-body and a monochromatic stellar spectrum in colliding-wind binaries]{reitberger14}. Assuming an isotropic distribution of electrons in the FF, the IC SED at each jet segment is computed as:
    \begin{equation}\label{LIC}
    \varepsilon' L_{\varepsilon'}^{\prime{\rm IC}}(l) = \varepsilon^{\prime 2} \int_0^\infty \frac{{\rm d}^2 N'}{{\rm d}\varepsilon' \ {\rm d}t'} N'(E',l) \ {\rm d}E' \ .
    \end{equation}

Once the luminosities are obtained in the FF, we transform them into what would be seen by the observer. For a stationary jet such as the one considered in this work, we have $\varepsilon L_\varepsilon = \varepsilon' L'_\varepsilon \delta_{\rm obs}^3/\gamma_{\rm j}$ \citep{sikora97}, where $\delta_{\rm obs} = [\gamma_{\rm j} (1 - \beta_{\rm j} \cos{\theta_{\rm obs}})]^{-1}$ is the Doppler factor between the emitter and the observer, the velocity of the former and the latter making an angle $\theta_{\rm obs}$ in the LF \citep[e.g.,][]{dermer02}. This transformation is done individually for each jet segment, as $\theta_{\rm obs}$, and thus $\delta_{\rm obs}$, varies from one to another. Photon energies as seen by the observer are $\varepsilon = \delta_{\rm obs} \varepsilon'$.

\subsection{Absorption mechanisms}\label{absorption}

Two main photon absorption processes are considered in this work: electron-positron pair creation from gamma rays interacting with stellar photons ($\gamma \gamma-$absorption, GGA), and the absorption of low-energy photons by the free ions present in the wind (free-free absorption, FFA). Synchrotron self-absorption is also computed and found to be negligible in comparison to FFA in this scenario, regardless of the adopted parameters. The absorption coefficients are calculated directly in the LF, as the emitted luminosities are previously transformed to what would be seen by the observer (see Sect.~\ref{radiation}).

The optical depth for GGA is computed as follows:
    \begin{equation}\label{taugg}
    \tau_{\gamma\gamma}(\varepsilon) = \int_0^d {\rm d}s \ [1-\cos{\theta_{\gamma\gamma}(s)}] \int_{\varepsilon_0^{\rm min}(\theta_{\gamma\gamma})}^\infty {\rm d}\varepsilon_0 \
    n(\varepsilon_0) \sigma_{\gamma\gamma}(\varepsilon,\varepsilon_0,s) \ ,
    \end{equation}
where $s$ parametrizes the gamma-ray photon trajectory along the line-of-sight, $d$ is the distance to the observer, $\theta_{\gamma\gamma}$ is the angle between the momentum of this photon and the stellar photons, $\varepsilon_0$ is the energy of the latter, which follows a black body distribution $n(\varepsilon_0)$, and  $\sigma_{\gamma\gamma}$ is the cross-section for GGA, obtained from Eq. (1) of \cite{gould67}. The lower integral limit $\varepsilon_0^{\rm min} = 2 m_{\rm e}^2 c^4/\varepsilon(1-\cos\theta_{\gamma\gamma})$ is the energy threshold for the creation of an electron-positron pair.

Regarding the FFA process, its absorption coefficient is given by \cite[e.g.,][]{rybicki86}:
    \begin{equation}\label{eq:alphaff}
    \alpha_\mathrm{ff} = \frac{4 e^6}{3 m_{\rm e} h c} \sqrt{\frac {2\pi}{3 k_{\rm B} m_{\rm e}}}
    Z^2 n_\mathrm{e} n_\mathrm{i} T^{-1/2} \nu^{-3} \left( 1 - {\rm e}^{-h\nu/k_{\rm B}T} \right) g_\mathrm{ff} \ ,
    \end{equation}
with $n_\mathrm{i,e}$ being the number density of ions/electrons in the wind, $T$ its temperature, $Z$ the mean atomic number (taken as 1), $\nu$ the photon frequency, and $g_\mathrm{ff}$ the average Gaunt factor, which can be estimated as $g_\mathrm{ff} \approx 9.77 [1 + 0.056\ln{(T^{3/2}/Z \nu)}]$ \citep{leitherer91}.

FFA is strongly dependent on the wind density (as $\propto n_\mathrm{w}^2$), and consequently would also be significantly affected by the presence of wind beaming (see Sect.~\ref{hydrodynamics}). Also, for $h \nu \ll k_{\rm B} T$ the absorption is $\propto \nu^{-2}$, so its effect is much larger for low radio frequencies. The FFA optical depth can be calculated as
    \begin{equation}\label{eq:tauff}
    \tau_{\rm ff}(\nu) = \int_0^d \alpha_{\rm ff}(\nu,r) \ {\rm d}s \ , 
    \end{equation}
where $r$ is the distance to the star. Given that the stellar wind has a density $n_\mathrm{w} \propto r^{-2}$ and that we are modeling a compact system in which the jet inner regions are close to the star, strong absorption is expected in radio.

In addition to the three aforementioned absorption processes, occultation of some parts of the jets by the star  is also taken into account \citep[e.g.,][]{khangulyan18}. Although unimportant for most system configurations, stellar occultation can have a moderate impact on the radiation output for high system inclinations ($i \gtrsim 60\degree$) when the CO is behind the star (see Sect.~\ref{orbit}).

\section{Results}\label{results}

We explore different values for three free parameters of our model: $\gamma_{\rm j}$, $\eta_B$, and $i$ (see Table~\ref{tab:parameters}). Moreover, we also study the orbital variability of the results by varying $\varphi$ between 0 and 1 (see Fig.~\ref{fig:sketch}). The energy losses and the particle energy distribution are only affected by $\gamma_{\rm j}$ and $\eta_B$, whereas the radiative outputs also depend on $i$ and $\varphi$. The observer is assumed to be always in a position such that the CO is closest to the observer for $\varphi = 0$ (inferior conjunction), and farthest from the observer for $\varphi = 0.5$ (superior conjunction). As we are considering a circular orbit, by studying a whole period we cover all the possible system configurations.

\subsection{Particle distribution}

\begin{figure}
    \centering
    \includegraphics[angle=270, width=\linewidth]{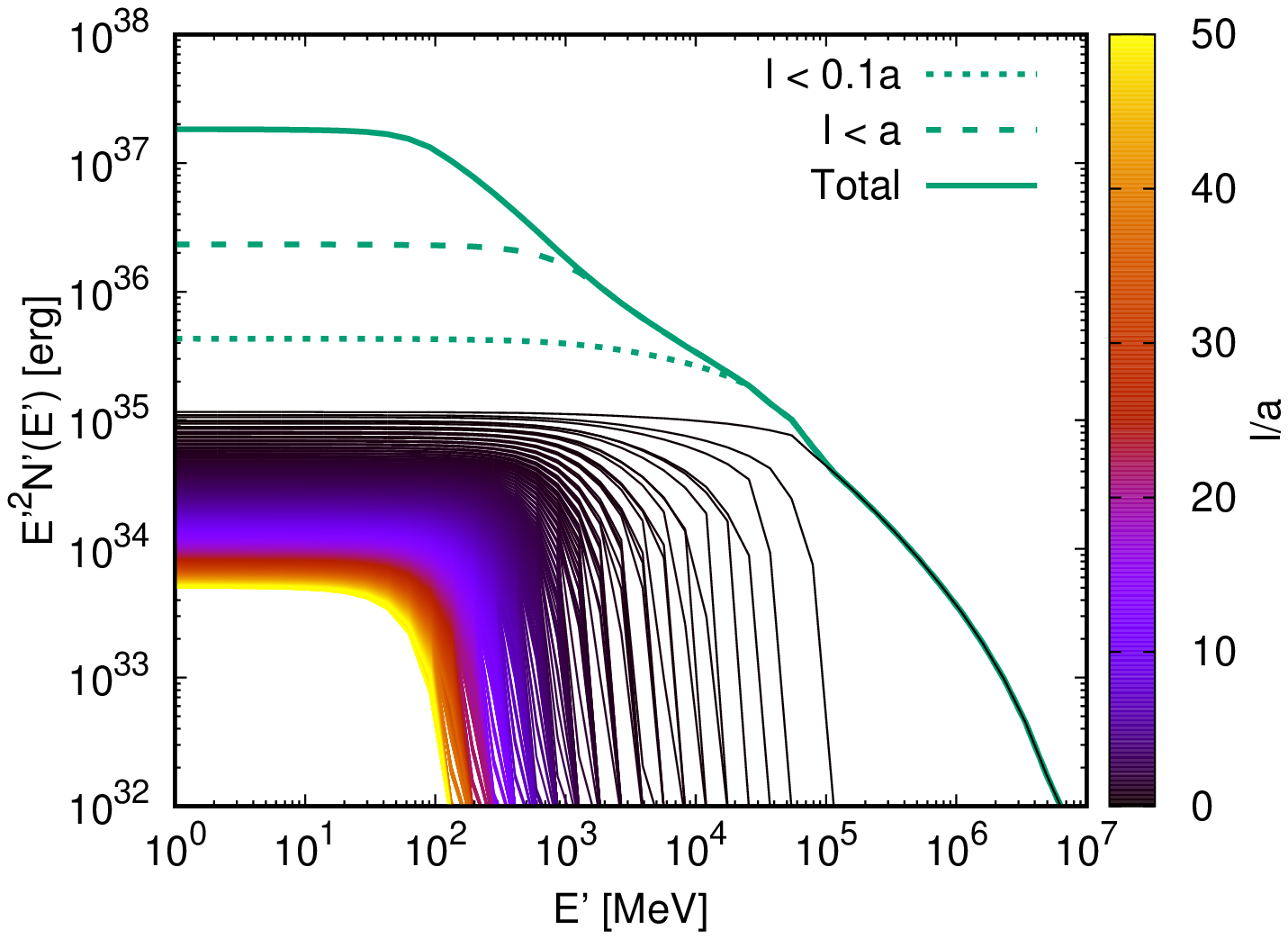} \\
    \vspace{3mm}
    \includegraphics[angle=270, width=\linewidth]{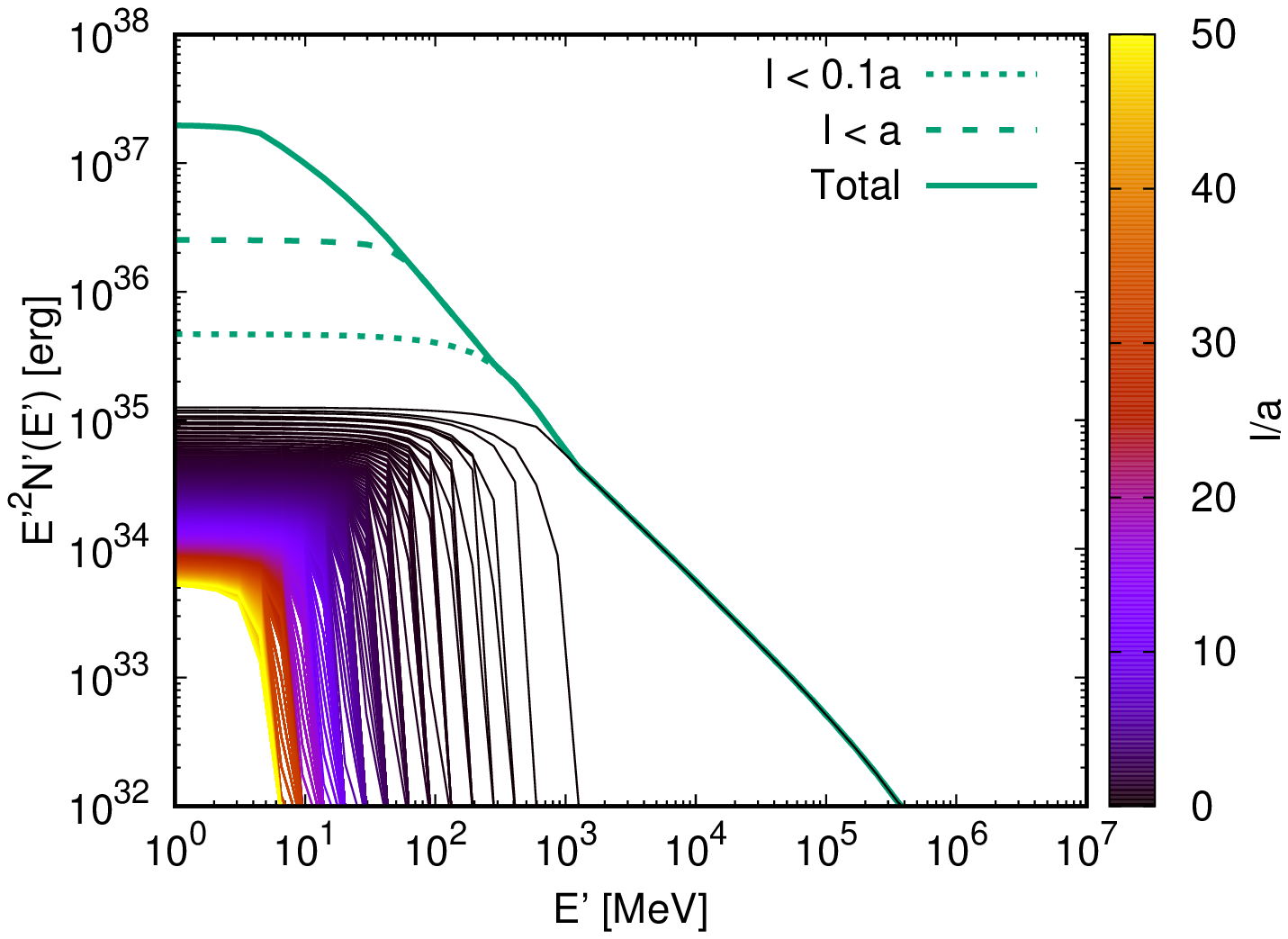}
    \caption{Energy distribution of the electrons along the jet up to $0.1a$ (dotted line), $a$ (dashed line), and the whole jet (solid line), for $\gamma_{\rm j} = 1.2$, $\eta_B = 10^{-2}$ (top panel), and $\eta_B = 1$ (bottom panel). The contribution of the individual jet segments is color-coded, with the color scale representing the position of each segment along the jet. Segments have a constant length of $\mathrm{d}l = 10^{11}$~cm $\approx 0.03 a$.}
    \label{fig:NE_fB}
\end{figure}

\begin{figure}
    \centering
    \includegraphics[angle=270, width=\linewidth]{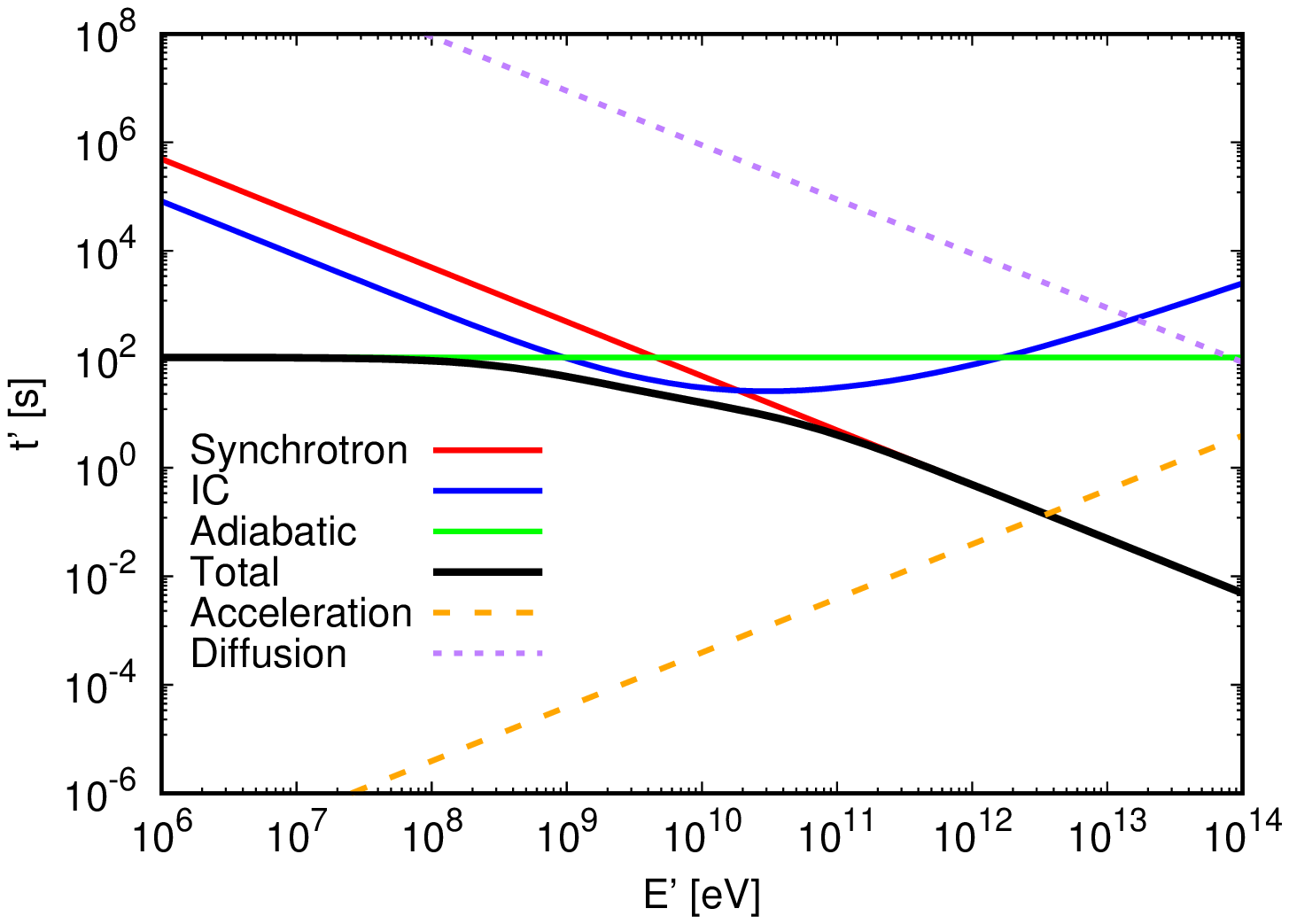}
    \caption{Characteristic timescales for the first jet segment for $\gamma_{\rm j} = 1.2$ and $\eta_B = 10^{-2}$. Cooling (solid lines), acceleration (dashed line), and diffusion (dotted line) processes are shown.}
    \label{fig:timescales}
\end{figure}

In Fig.~\ref{fig:NE_fB}, we show the electron energy distribution of each jet segment for $\gamma_{\rm j} = 1.2$ and $\eta_B = 10^{-2}$ and $1$ (which yield $B'_0 = 28.2$~G and $282$~G, respectively), as well as the total electron energy distribution up to three different lengths along the jet. The relevant timescales for the first jet segment, where relativistic electrons are injected, are shown in Fig.~\ref{fig:timescales} for $\gamma_{\rm j} = 1.2$ and $\eta_B = 10^{-2}$. Both the electron distribution and the timescales associated to it are the same for the jet and the counter-jet. For $\eta_B=1$, synchrotron is the dominant cooling mechanism for electrons with $E' \gtrsim 50$~MeV, and those with $E' \gtrsim 1$~GeV cool down already within the first segment. At lower electron energies the cooling is dominated by the adiabatic expansion of the jet, which allows most of the particles below $\sim 50$~MeV to reach distances outside the binary system. For $\eta_B = 10^{-2}$, the synchrotron dominance only happens above $\sim 30$~GeV, and there is also a significant contribution of IC losses for $1$~GeV $\lesssim E' \lesssim 70$~GeV, as seen in Fig.~\ref{fig:timescales}. However, this contribution is only relevant for the inner $l \sim 10^{13}$~cm of the jet, where the stellar radiation field is strong; farther away adiabatic losses also dominate in this energy range.

\subsection{Spectral energy distribution}

\begin{figure}
    \centering
    \includegraphics[angle=270, width=\linewidth]{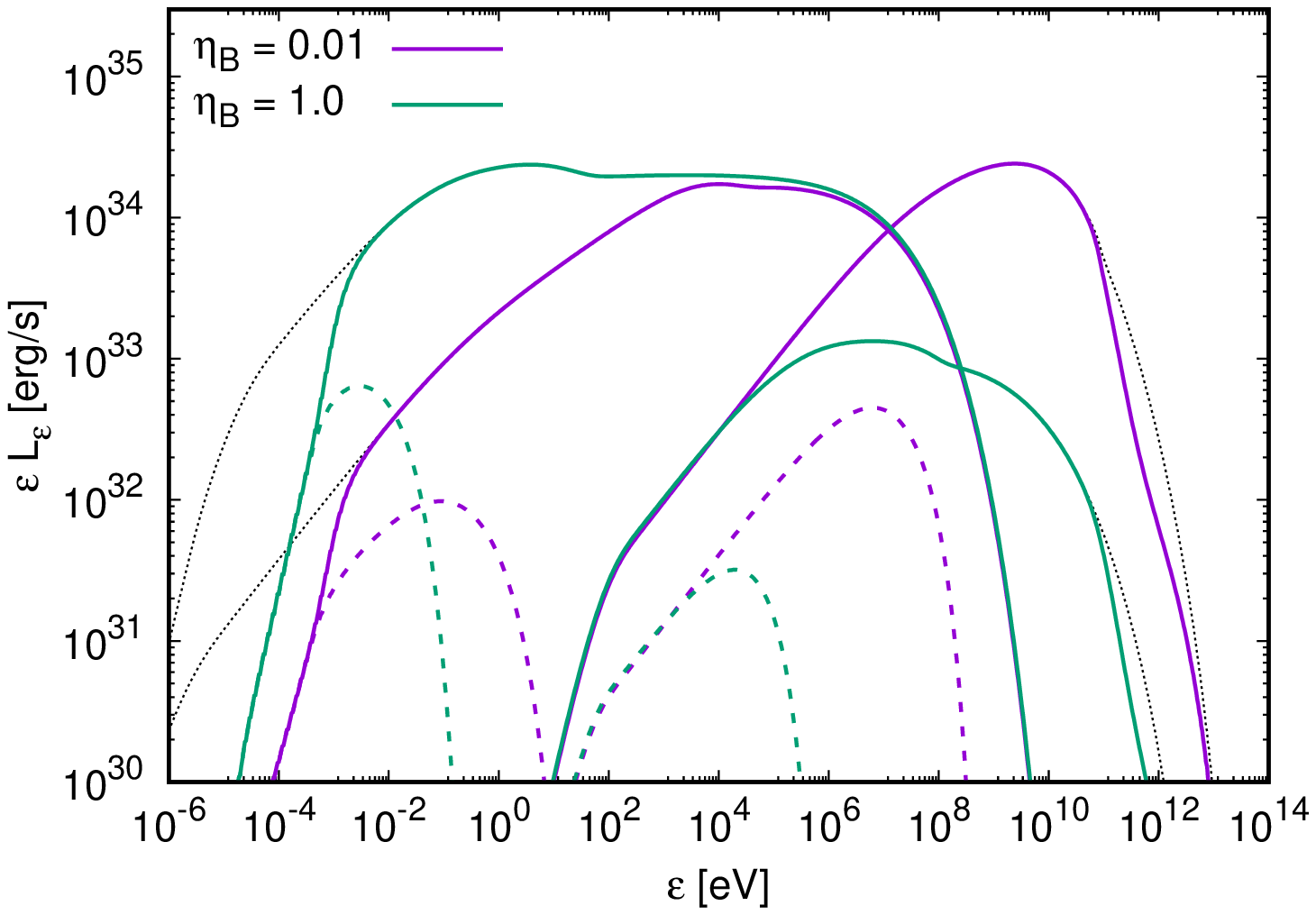}
    \caption{Total (jet + counter-jet) synchrotron and IC SEDs for $\gamma_{\rm j} = 1.2$, $i = 30\degree$, $\varphi = 0.25$, and $\eta_B = 10^{-2}$ (purple lines) and $1$ (green lines). Dashed lines show the contribution of the jets beyond a length $a$, and black dotted lines represent the unabsorbed emission.}
    \label{fig:SED_total}
\end{figure}

The SEDs in this section are computed using $\varphi = 0.25$ as a representative situation. Figure~\ref{fig:SED_total} shows the combined SED (as seen by the observer) of the jet and the counter-jet for $\gamma_{\rm j} = 1.2$, $i = 30\degree$, and $\eta_B = 10^{-2}$ and $1$. The contribution of the regions of the jets further from to the CO is also depicted. The bump at low energies of the synchrotron SED for $\eta_B = 1$ is a numerical artifact without physical meaning caused by how the first jet segment is treated when computing the particle distribution for intense energy losses \citepalias[see][a bump is also hinted in the electron energy distribution in Fig.~\ref{fig:NE_fB}]{molina18}. The $\eta_B$-value has a strong influence on the synchrotron and IC luminosities, as expected from its relation to $\eta_B^\star$. The SED is totally dominated by the emission at the binary system scales, except for the low-energy end of the synchrotron component, in which the radiation output comes mainly from the more external regions given the high level of FFA close to the star. This same absorption process is the responsible for the lack of radio emission in the spectrum. GGA is significant even for small system inclinations, despite a low $i$ not favouring this absorption process.

\begin{figure}
    \centering
    \includegraphics[angle=270, width=\linewidth]{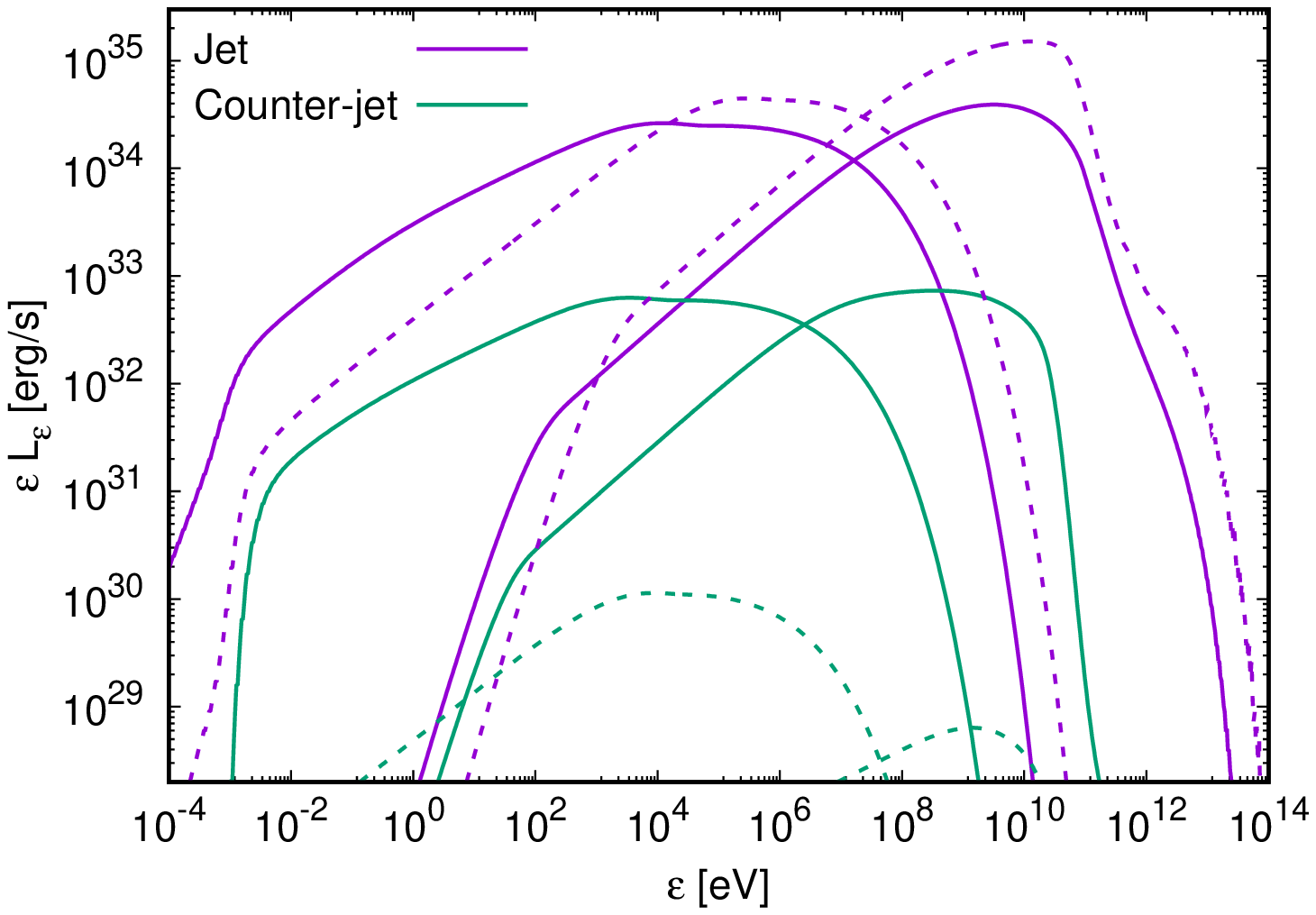} \\
    \vspace{3mm}
    \includegraphics[angle=270, width=\linewidth]{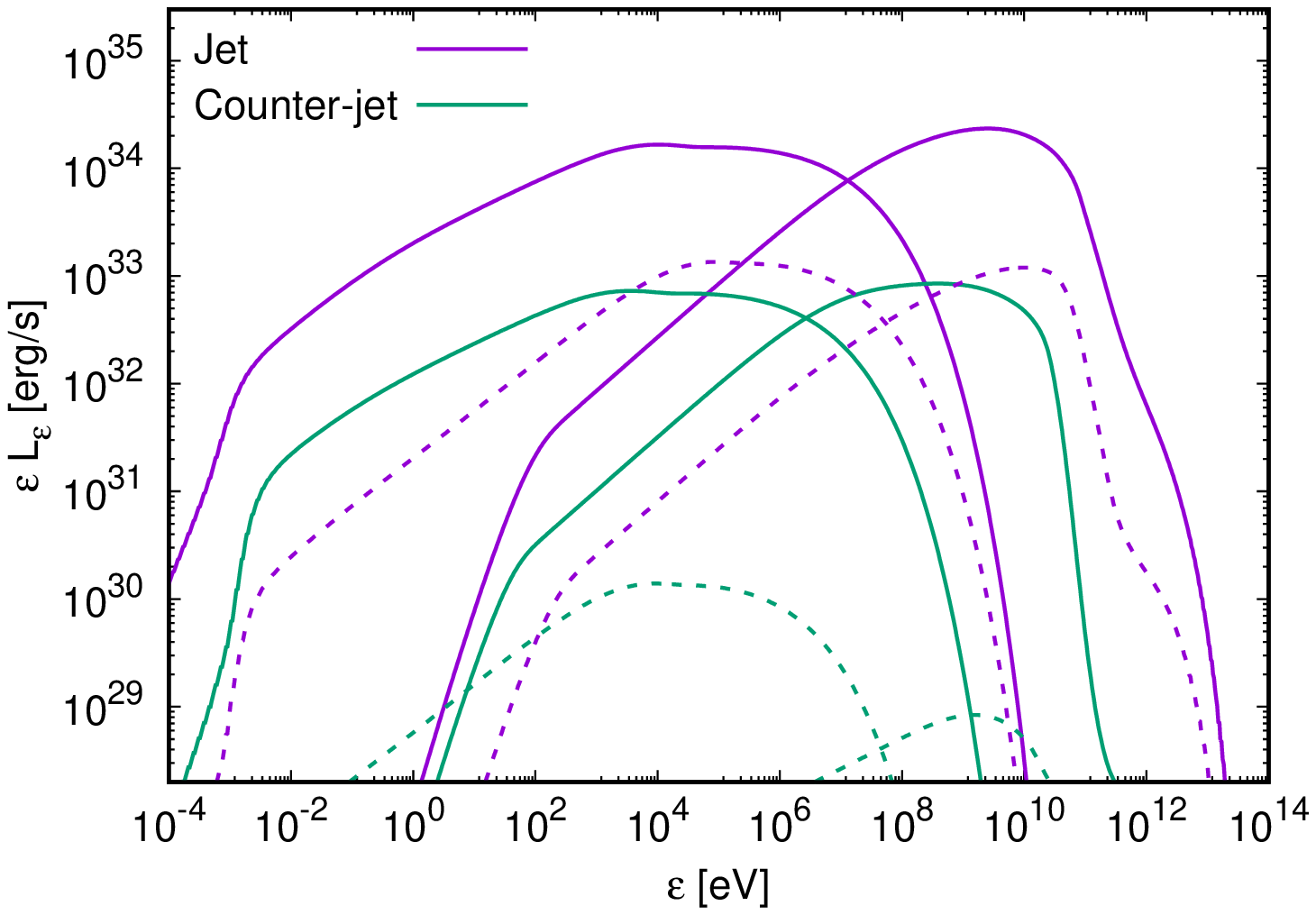} \\
    \vspace{3mm}
    \includegraphics[angle=270, width=\linewidth]{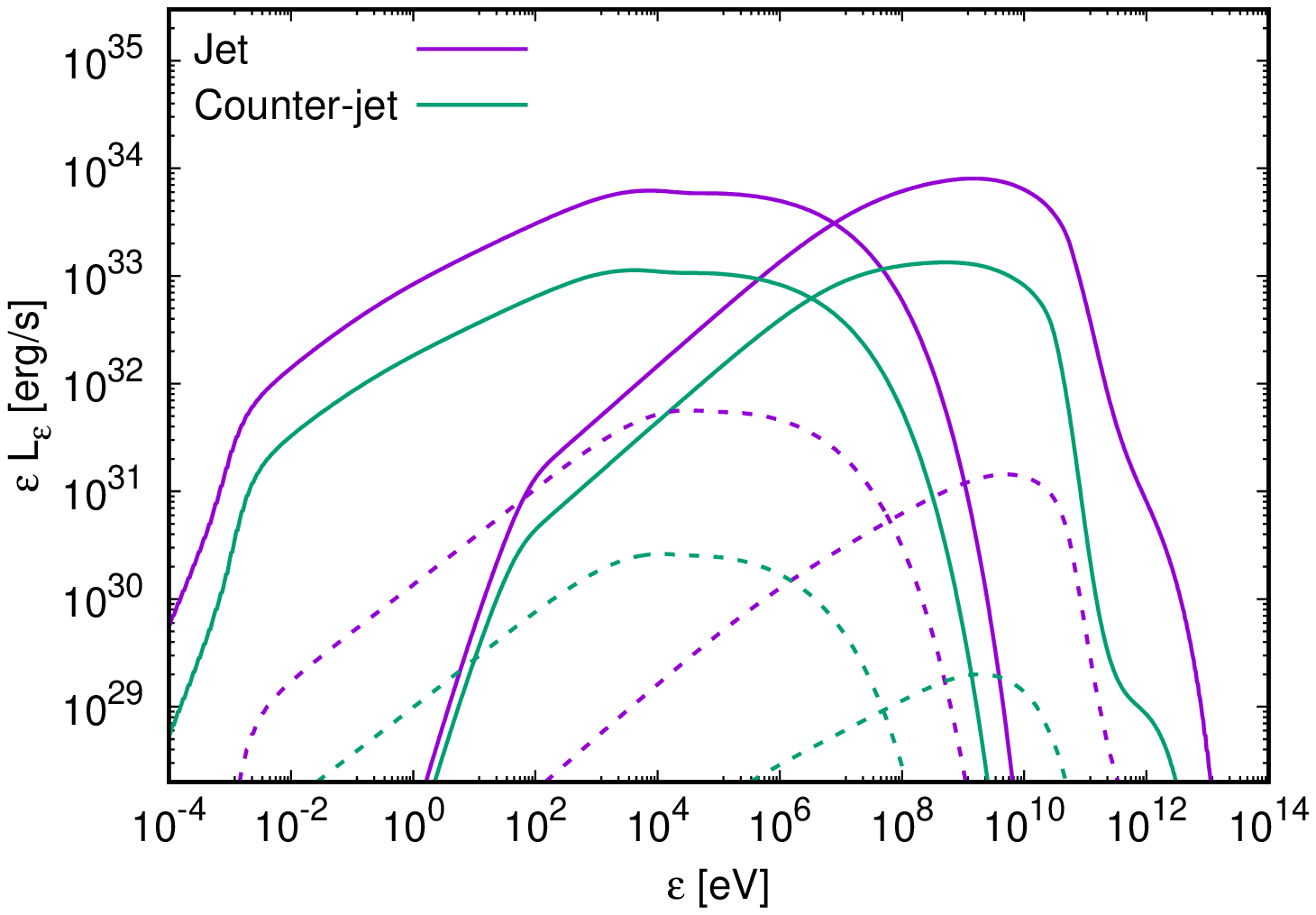}
    \caption{SEDs of the jet (purple lines) and the counter-jet (green lines) for $\eta_B = 10^{-2}$, $\varphi = 0.25$, $i = 0$ (top) $30\degree$ (middle) and $60\degree$ (bottom), and $\gamma_{\rm j} = 1.2$ (solid lines) and $3$ (dashed lines).}
    \label{fig:SED_2jets}
\end{figure}

\begin{figure}
    \centering
    \includegraphics[angle=270, width=\linewidth]{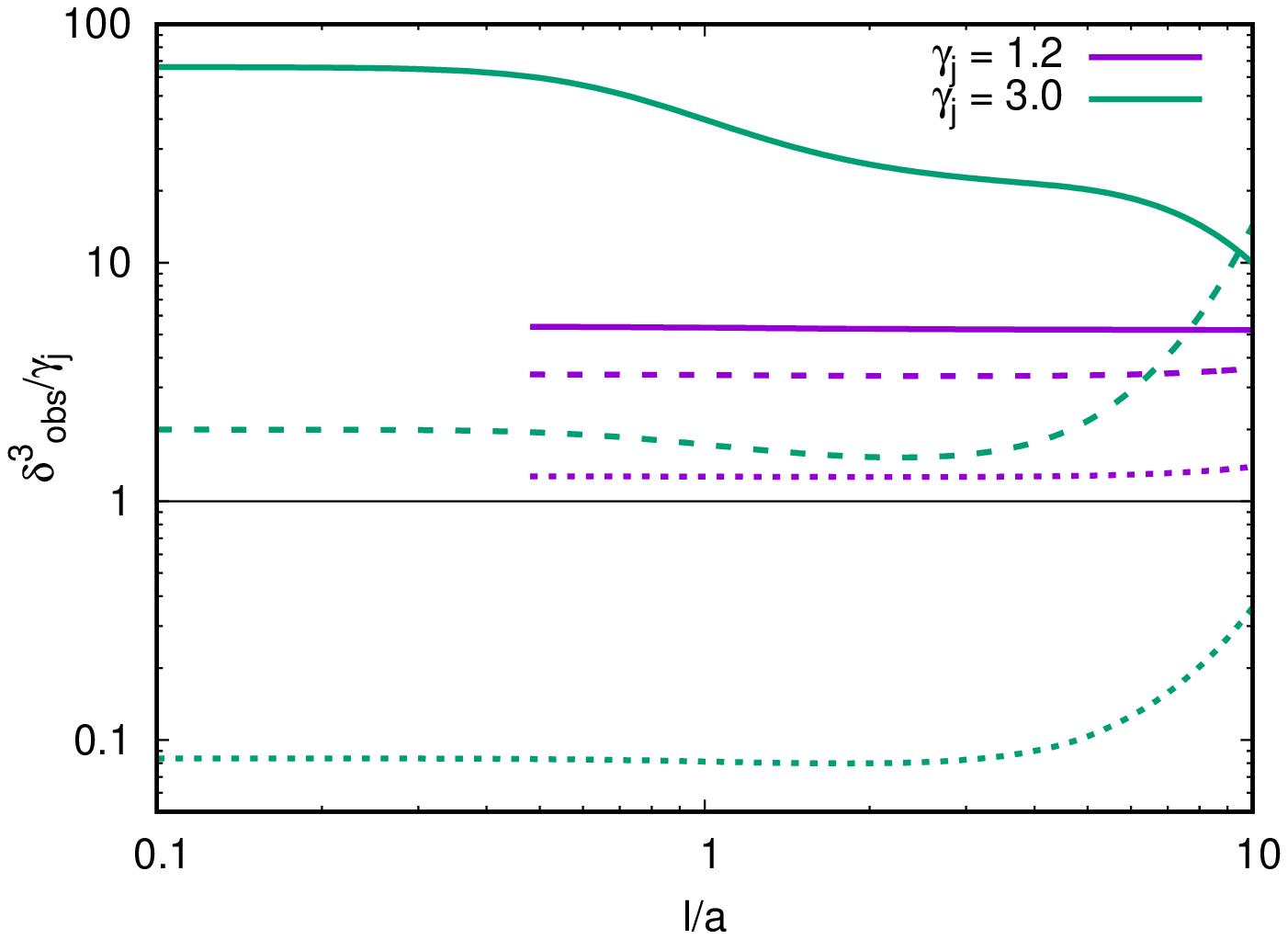}
    \caption{Doppler boosting factor as a function of jet position for $\varphi=0.25$, $\gamma_{\rm j}= 1.2$ and $3$, and $i = 0$ (solid lines), $i = 30\degree$ (dashed lines), and $i = 60\degree$ (dotted lines). The black solid line at $\delta_{\rm obs}^3/\gamma_{\rm j}~=~1$ sets the border between boosting (> 1) and deboosting (< 1). Only the values up to $l = 10a$ are shown.}
    \label{fig:boosting}
\end{figure}

The separate jet and counter-jet SEDs are presented in Fig.~\ref{fig:SED_2jets} for $i = 0\degree$, $30\degree$ and $60\degree$, $\eta_B = 10^{-2}$, and $\gamma_{\rm j} = 1.2$ and $3$. The jet is responsible for most of the emission even for a high inclination, which reduces the effect of Doppler boosting. Figure~\ref{fig:boosting} shows the evolution along the jet of the Doppler boosting factor, $\delta_{\rm obs}^3/\gamma_{\rm j}$, for $\gamma_{\rm j} = 1.2$ and $3$, and $i = 0\degree$, $30\degree$ and $60\degree$. Results are only shown up to $l = 10a$, as the contribution of the jets to the SED beyond this length is negligible. For $\gamma_{\rm j} = 1.2$, $\delta_{\rm obs}^3/\gamma_{\rm j}$ remains almost constant because $\theta_{\rm obs}$ does so, given the small jet deviation with respect to a straight trajectory at the scales shown in the plot (see Fig.~\ref{fig:trajectory}). For this $\gamma_{\rm j}-$value, the emission is boosted with respect to the FF as long as $i \lesssim 60\degree$. For $\gamma_{\rm j} = 3$ the variation of the Doppler boosting factor along the jet becomes more important due to its helical shape, and the emission is already deboosted for $i \gtrsim 30\degree$. We note that a higher $\gamma_{\rm j}$ implies more boosting only when $i \sim 0$. This can further be seen in Fig.~\ref{fig:SED_2jets}, where only for the case of $i = 0$ the SED for $\gamma_{\rm j} = 3$ becomes comparable to that for $\gamma_{\rm j} = 1.2$.

The intrinsic jet emission is also highly influenced by the value of $\gamma_{\rm j}$ as the latter affects the magnetic and photon fields in the FF. In terms of particle cooling, the relative importance of each process varies with $\gamma_{\rm j}$. Synchrotron losses become less efficient for increasing Lorentz factor, as $t'_{\rm syn} \propto B^{\prime -2} \propto \gamma_{\rm j}^2$, whereas adiabatic losses become more dominant, as $t'_{\rm ad} \propto \gamma_{\rm j}^{-1}$. The dependence of IC cooling with $\gamma_{\rm j}$ is more complex \citep[see][]{khangulyan14}, but for the cases considered in this work the variation of $t'_{\rm IC}$ is small compared to that of $t'_{\rm syn}$ and $t'_{\rm ad}$. In general, the combination of the effects of $\gamma_{\rm j}$ on the intrinsic and observed radiative outputs results in a decrease of the detected emission when $\gamma_{\rm j}$ increases, except for small enough values of $i$.

\subsection{Orbital variability}\label{orbit}

\begin{figure}
    \centering
    \includegraphics[angle=270, width=\linewidth]{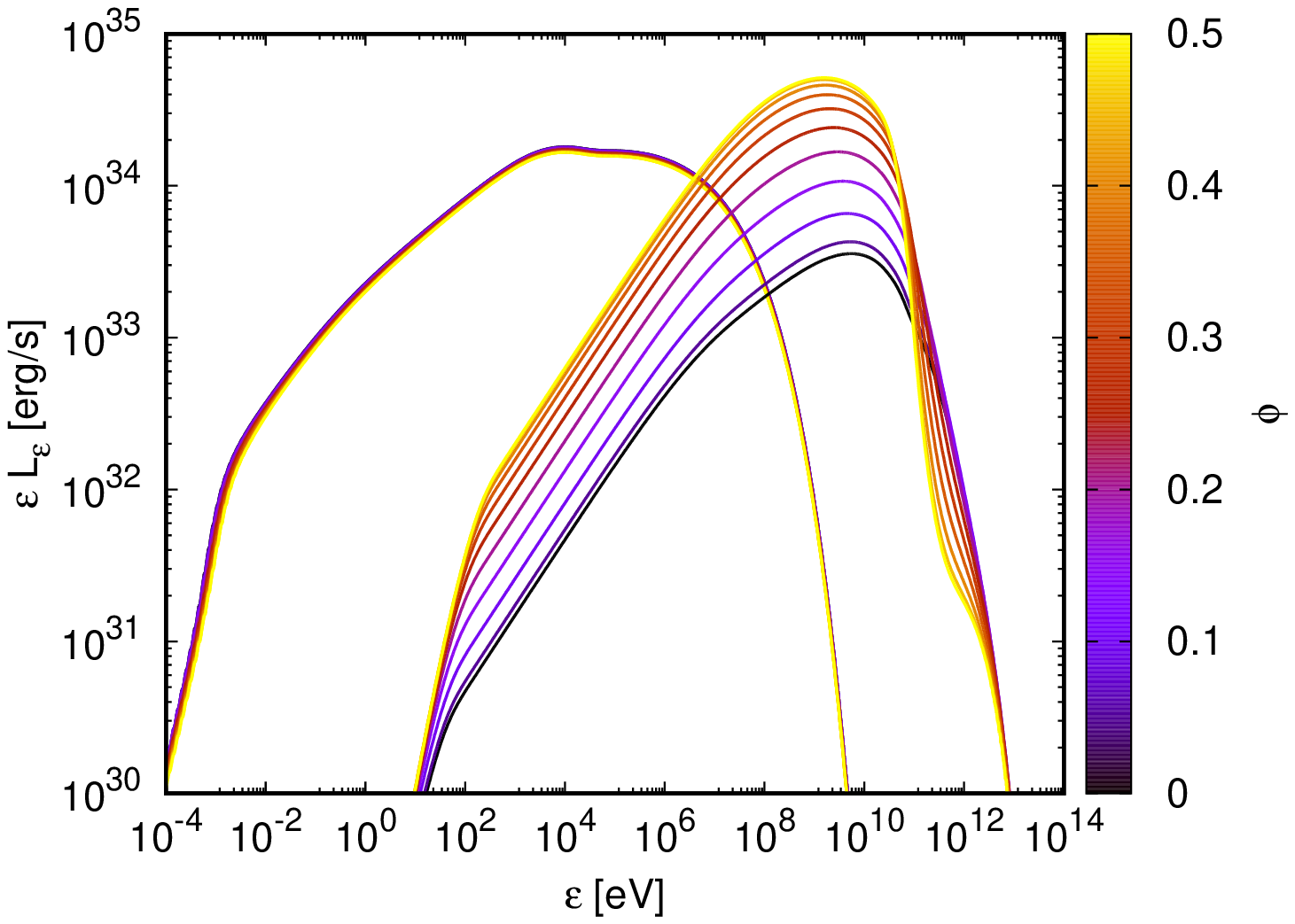} \\
    \vspace{3mm}
    \includegraphics[angle=270, width=\linewidth]{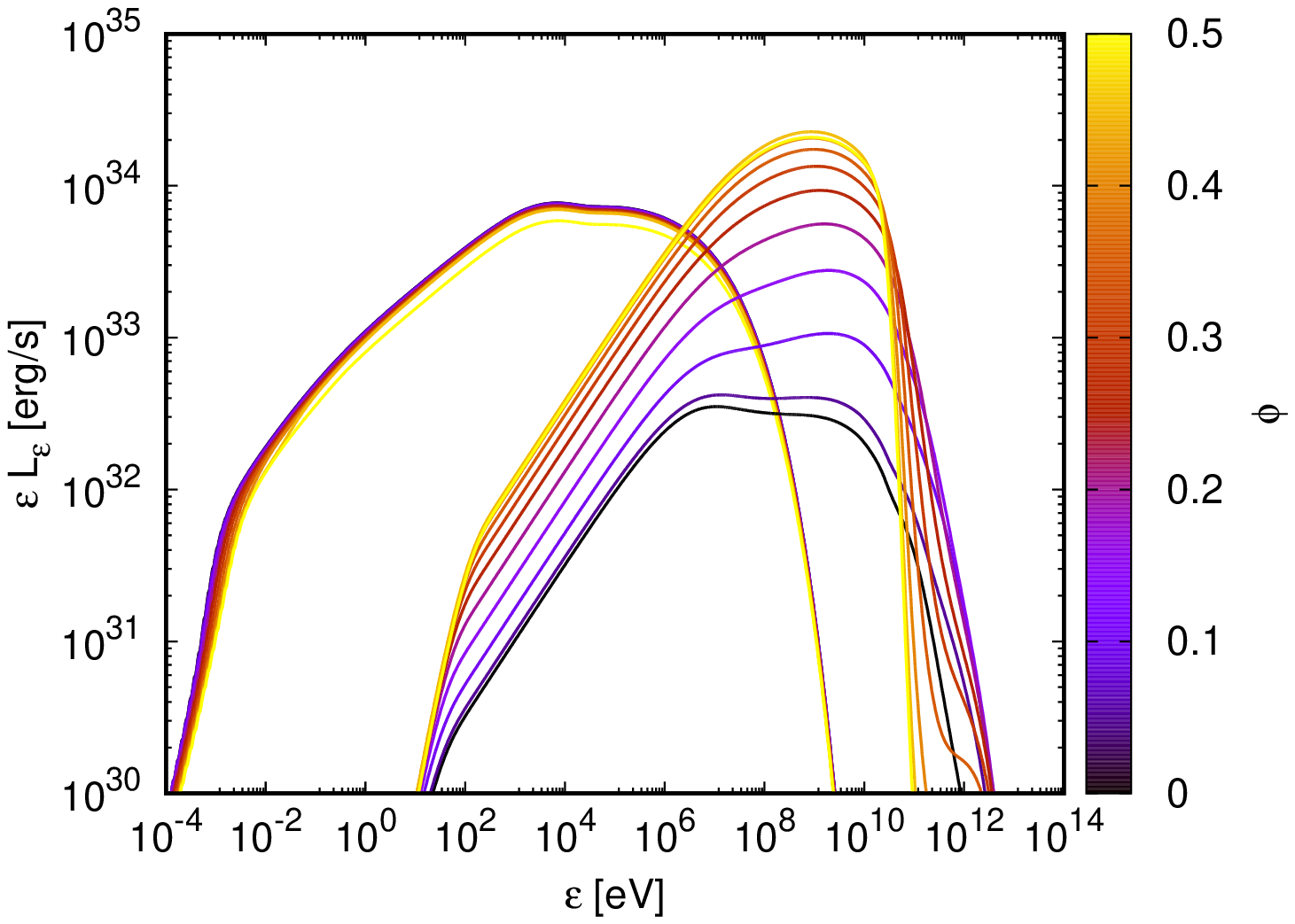}
    \caption{Total synchrotron and IC SEDs for $\gamma_{\rm j} = 1.2$, $\eta_B = 10^{-2}$, $i = 30\degree$ (top) and $60\degree$ (bottom), and different orbital phases between 0 and 0.5. SEDs for $0.5 < \varphi < 1$ are not represented because they almost overlap with those shown.}
    \label{fig:SED_orb_1.2}
\end{figure}

\begin{figure}
    \centering
    \includegraphics[angle=270, width=\linewidth]{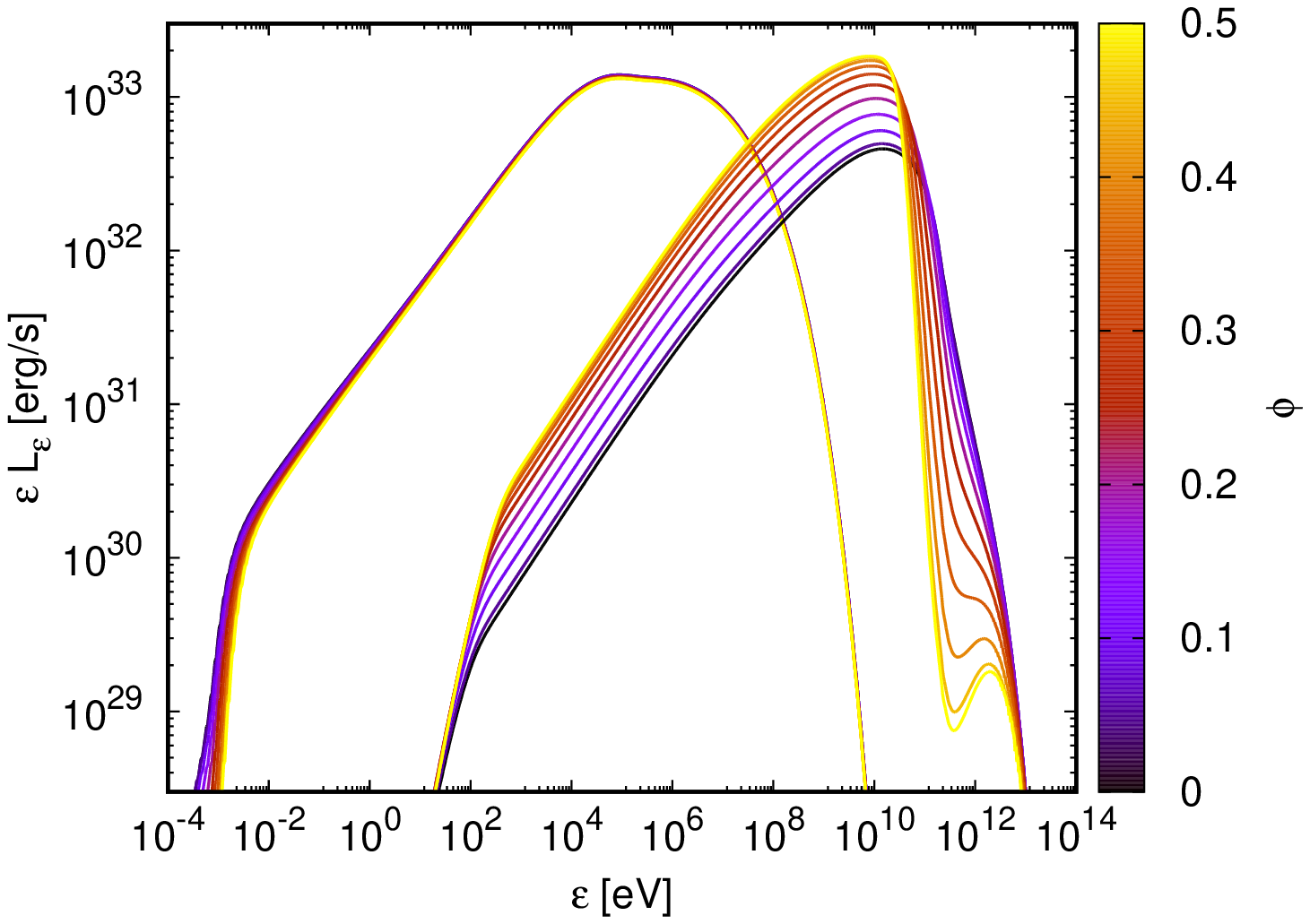} \\
    \vspace{3mm}
    \includegraphics[angle=270, width=\linewidth]{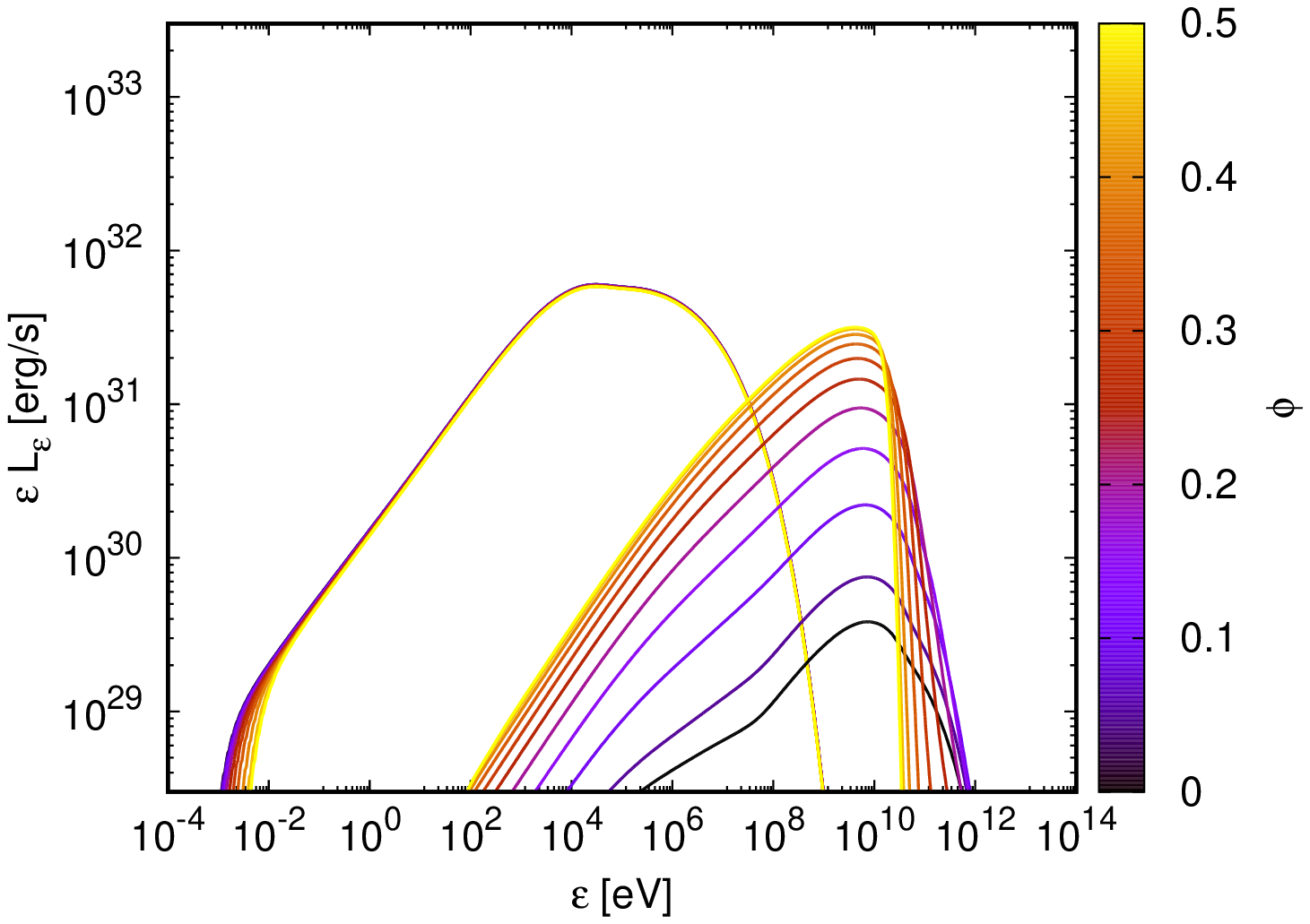}
    \caption{Same as in Fig.~\ref{fig:SED_orb_1.2}, but for $\gamma_{\rm j} = 3.0$. Note the change in the $y$-axis scale.}
    \label{fig:SED_orb_3.0}
\end{figure}

Figures~\ref{fig:SED_orb_1.2} and \ref{fig:SED_orb_3.0} show the total SEDs for $\eta_B = 10^{-2}$, $i = 30\degree$ and $60\degree$, and different values of $\varphi \in [0,0.5]$, $\gamma_{\rm j} = 1.2$ and $3$, respectively. Emission is almost symmetric for the remaining orbit, that is $\varphi \in (0.5,1)$, despite the helical trajectory of the jets. The difference, thus, being only of a few percent, is not shown in the plot. Synchrotron emission at $\varepsilon \gtrsim 10^{-2}$~eV remains almost constant over the whole orbit, with slight variations corresponding to small changes in the $\delta_{\rm obs}$ in the inner jet regions. At lower energies, especially for $i= 60\degree$, the difference becomes a bit more noticeable due to FFA becoming stronger as the CO approaches the superior conjunction ($\varphi = 0.5$). Both IC emission and GGA are strongly affected by the orbital phase and steadily increase as $\varphi$ goes from $0$ to $0.5$. This makes the IC radiation output at most photon energies increase towards the superior conjunction, with the exception of the highest gamma-ray energies ($\varepsilon \gtrsim 100$~GeV), for which the emission of the phases closer to the inferior conjunction is larger due to less efficient GGA. This effect is more pronounced for high inclinations, as the strength of GGA increases with $i$ close to the superior conjunction.

\begin{figure}
    \centering
    \includegraphics[angle=270, width=\linewidth]{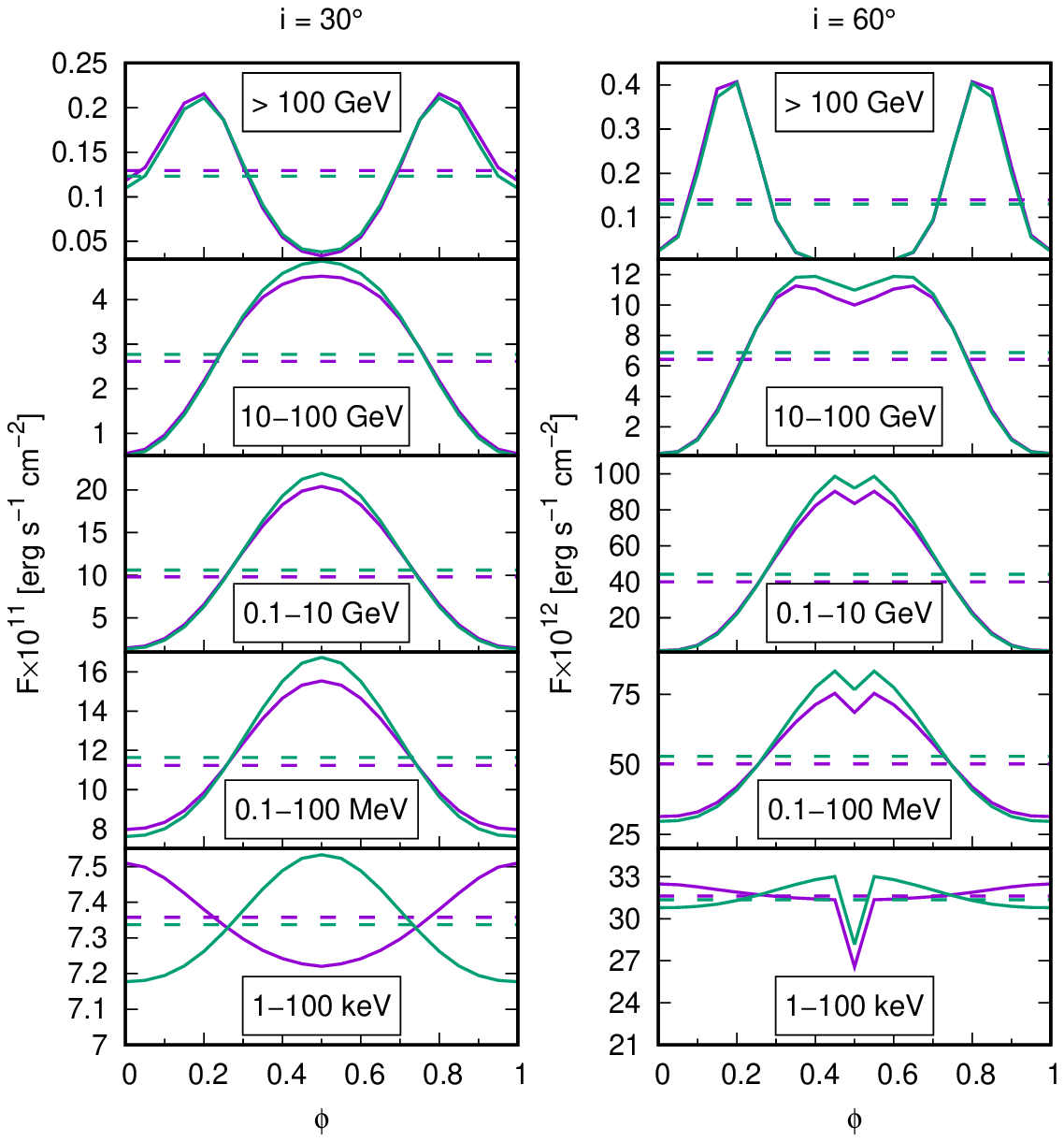}
    \caption{Integrated light curves for $\gamma_{\rm j} = 1.2$, $\eta_B = 10^{-2}$, $i=30\degree$ and $60\degree$, and different photon energy ranges. Purple lines represent the values for bent jets such as those in Fig.~\ref{fig:trajectory}, whereas green lines represent the results for straight jets. Dashed lines show the orbit-averaged flux values. Note the change in the flux normalization for each $i$.}
    \label{fig:fluxes_1.2}
\end{figure}

\begin{figure}
    \centering
    \includegraphics[angle=270, width=\linewidth]{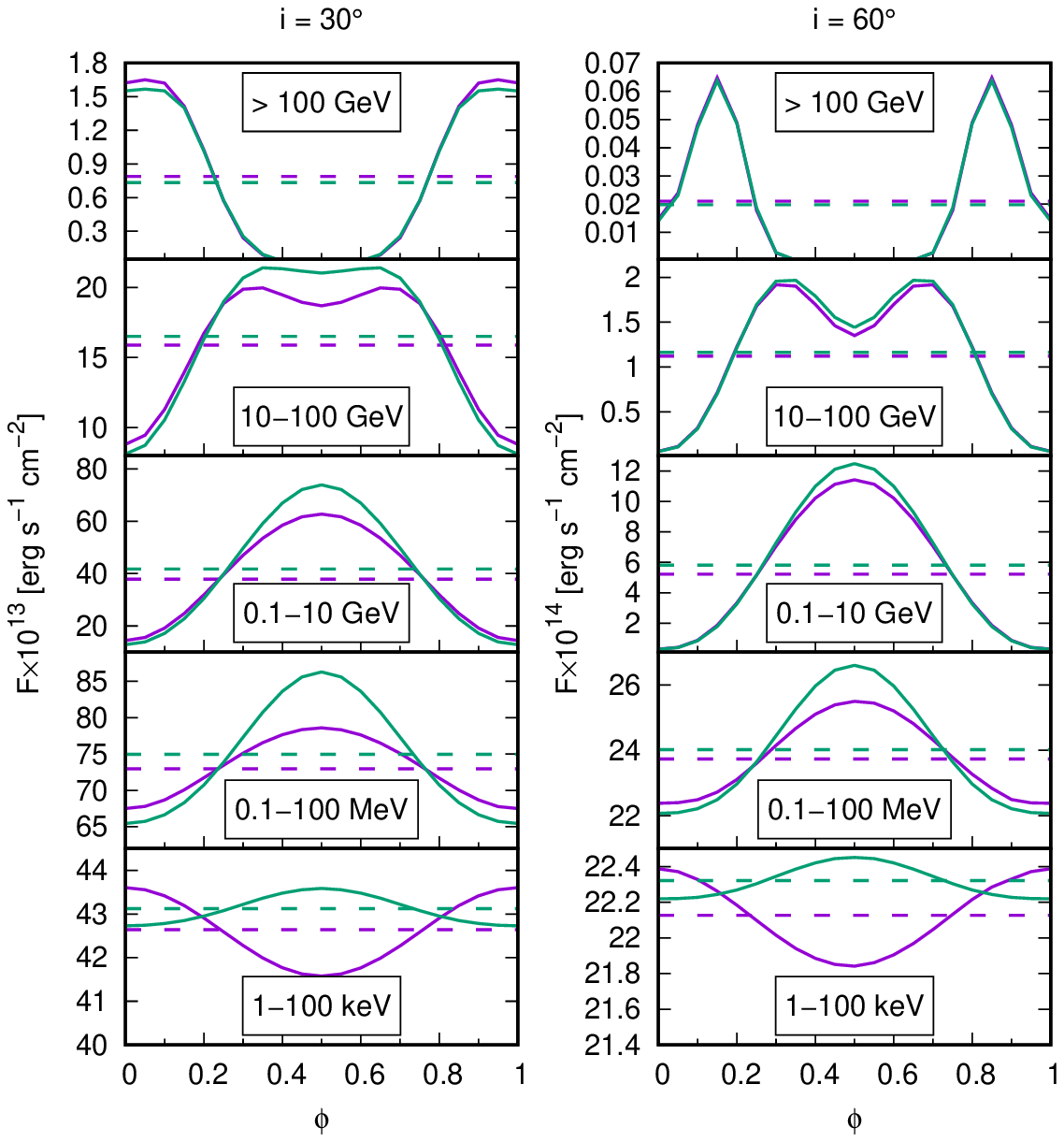}
    \caption{Same as in Fig.~\ref{fig:fluxes_1.2}, but for $\gamma_{\rm j} = 3$.}
    \label{fig:fluxes_3.0}
\end{figure}

Figures~\ref{fig:fluxes_1.2} and \ref{fig:fluxes_3.0} show the light curves at different energy bands for $\gamma_{\rm j} = 1.2$ and $3$, respectively. Two inclinations, $i = 30\degree$ and $60\degree$, are studied for $\eta_B = 10^{-2}$. Light curves for straight (unbent) jets moving perpendicular to the orbital plane are also shown for comparison, although we note that this is not a realistic case given the adopted wind and jet parameters. The bumps seen at $\varphi = 0.5$ and $\varepsilon \leq 10$~GeV are caused by photon occultation by the star. Orbital modulation for $\varepsilon \ge 100$~MeV becomes more important the higher the system inclination is. As an example, in the case of $\gamma_{\rm j} = 3$ (Fig.~\ref{fig:fluxes_3.0}) and $10$~GeV $\leq \varepsilon \leq 100$~GeV, for $i=30\degree$ the flux changes only by a factor of $\sim 3$ along the orbit, whereas for $i=60\degree$ it changes by a factor of $\sim 20$. This effect is not observed for $\varepsilon < 100$~MeV, as the weakly $\varphi$-dependent synchrotron emission dominates at these energies.

\begin{figure}
    \centering
    \includegraphics[angle=270, width=\linewidth]{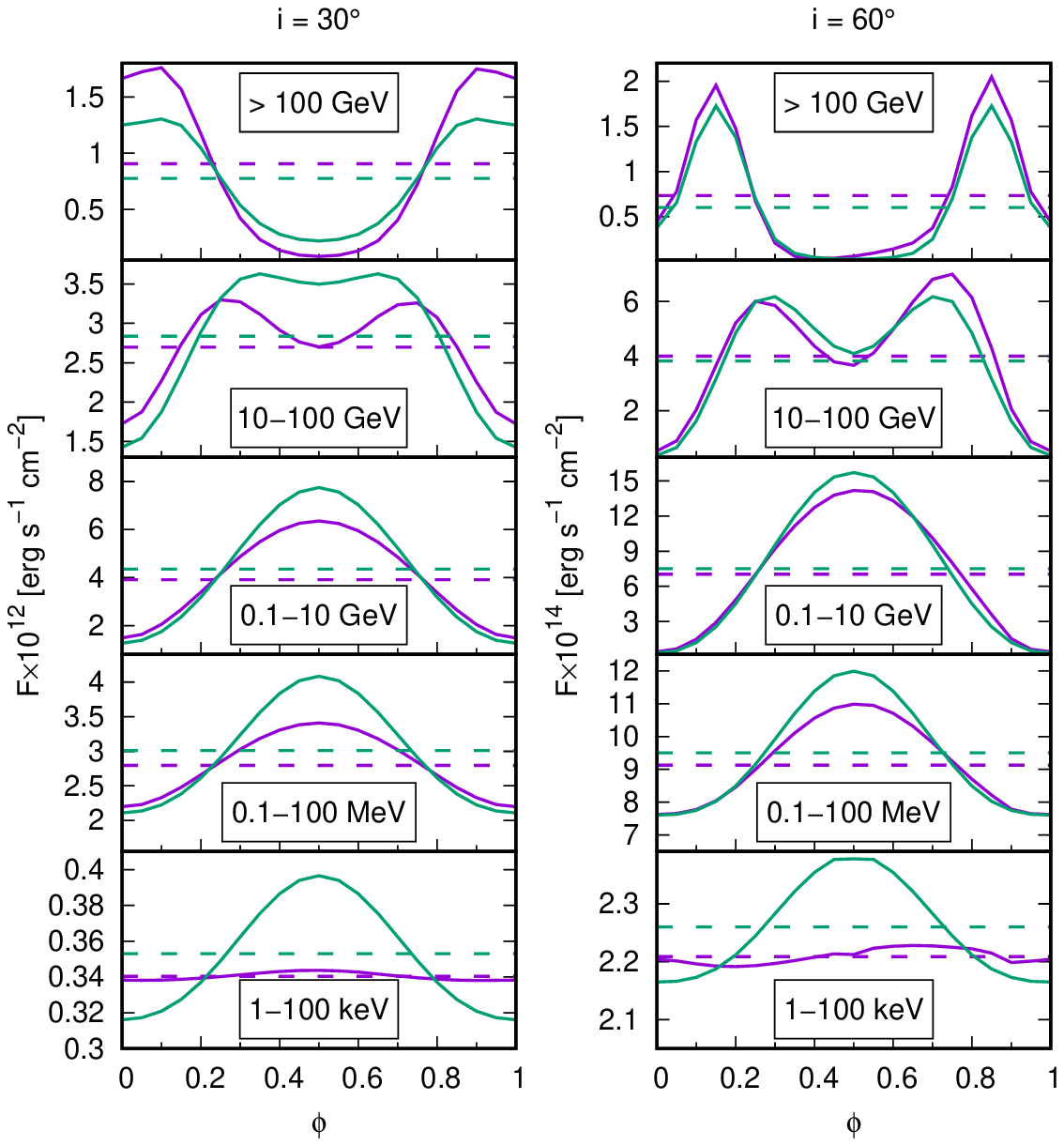}
    \caption{Same as in Fig.~\ref{fig:fluxes_1.2}, but for $\gamma_{\rm j} = 3$ and $\eta_B = 10^{-4}$.}
    \label{fig:fluxes_3.0_fB1e-4}
\end{figure}

The fluxes for bent and straight jets are similar between them, with differences of at most $\sim 15\%$ for $\eta_B = 10^{-2}$. The situation is the same when one considers the orbit-averaged values. This is explained by the strong concentration of the emission at the jet inner regions ($\lesssim a$; see Fig.~\ref{fig:SED_total}), where the difference in the shape of straight and helical jets is small. For $\gamma_{\rm j} = 1.2$, this difference remains small even for $l \gg a$. If one considers a lower magnetic field, however, the situation changes. Figure~\ref{fig:fluxes_3.0_fB1e-4} shows the light curves for $\gamma_{\rm j} = 3$ and $\eta_B = 10^{-4}$ ($B'_0 = 4.04$~G). As the energy losses close to the CO are less intense, IC emission is more significant for $l > a$ and the light curves have distinct features due to the helical jet trajectory. On the one hand, the difference in the fluxes with respect to straight jets is higher, up to $\sim 40\%$ for $\varepsilon > 10$~GeV. On the other hand, asymmetries in the light curves with respect to $\varphi = 0.5$ arise for $i = 60\degree$.

\section{Summary and discussion}\label{discussion}

We have extended the study performed in \citetalias{molina18} to the inner regions of HMMQ jets in order to assess the role of the wind-jet interaction at the binary system scales. On those scales, the jet is likely to be more relativistic than further away from the binary \citepalias[i.e., the scales studied in][]{molina18}. We have shown that the stellar wind plus orbital motion can have a significant effect on the jet radiative output. As a result of this wind-jet interaction, the jets get a helical shape that affects angle-dependent processes such as IC scattering, GGA, and relativistic Doppler boosting. Most of the jet emission is produced at distances below a few orbital separations. If one compares the present results to those of \citetalias{molina18}, one finds that the predicted fluxes are similar, although $L_{\rm NT}$ is here an order of magnitude higher, meaning more efficient acceleration is assumed in the strong, asymmetric jet recollimation shock than in the shock at larger scales considered in \citetalias{molina18}. The radio emission is severely absorbed in the stellar wind via FFA. This prevents us to use this emission to trace the helical jet structure, as it was done in \citetalias{molina18}. Additionally, the small scales considered in this work would require a very high ($\mu$-arcsecond) angular resolution to address such a study. The high proximity to the star also makes IC emission and GGA more susceptible to change along the orbit, which results in a much stronger orbital modulation than in \citetalias{molina18}, especially for high-energy gamma rays with $\varepsilon \gtrsim 100$~MeV. Evidence of such a modulation has been found for Cygnus~X-3 \citep{zdziarski18}, and hints of such an effect have also been observed in Cygnus~X-1 \citep{zanin16,zdziarski17}.

Including the helical shape of the jets has a moderate impact on the predicted light curves with respect to the straight jet case, introducing a variation not higher than $\sim 40\%$, with this value depending on the strength of the magnetic field. At the energies at which GGA plays an important role ($\varepsilon \gtrsim 10$~GeV), the emission from helical jets is consistently more absorbed close to the superior conjunction due to the additional distance that the gamma-rays have to travel embedded in the stellar photon field. Nonetheless, this feature must be taken with caution, as a similar effect could be seen for straight jets with acceleration at $z < z_{\rm rec}$, not associated with the recollimation shock. Moreover, helical jets may have stronger energy dissipation, and thereby particle acceleration, than straight jets at $z > z_{\rm rec}$ (not considered here), which could counterbalance also this effect. For small magnetic fields, the asymmetry in the light curves of jets with a helical shape could hint at the presence of this kind of jet structure in HMMQs. This asymmetries were also obtained by \cite{dubus10} when studying the effect of jet precession in the gamma-ray light curve of Cygnus~X-3.

The parameters $\eta_{\rm NT}$ and $\eta_{\rm acc}$ used in our model are very difficult to constrain both observationally and from first principles, and thus we have only taken representative values for them. In the case of $\eta_{\rm NT}$, the values of the luminosities and fluxes are easily scalable as $\propto \eta_{\rm NT}$. The value of $\eta_{\rm acc}$ affects the maximum energy that the electrons can attain, and consequently the peak position of the synchrotron and IC components in the SED. A significantly lower value of $\eta_{\rm acc}$ would not allow synchrotron (IC) photons to reach X-rays (gamma rays above $100$~GeV), whereas the GeV emission would remain relatively unaffected. As noted, it is also possible that electrons are accelerated in a continuous region extending beyond $\overrightarrow{r_0}$. This would modify the particle distribution in such a way that relativistic electrons could propagate further downstream along the jet without cooling down within the first jet segment. As synchrotron losses would not be so dominant further from the CO, this would result in an increase in the emitted gamma-ray luminosity and a decrease in the X-ray luminosity. In addition, the maximum electron and emitted photon energies could be somewhat higher if not limited by synchrotron cooling but by the accelerator size \citep[e.g.,][]{khangulyan08}. In the present work, for simplicity, we neglect electron acceleration beyond $z_{\rm rec}$, but this possibility should be considered in future studies.

Another free parameter in our model is related to the magnetic field strength, $\eta_B$. As shown in Fig.~\ref{fig:SED_total}, this parameter has a strong impact on the produced fluxes at all photon energies, and it also affects the degree of concentration of the emission close to the CO: the higher $\eta_B$ is, the more concentrated the emission is. The position of the synchrotron high-energy cutoff is rather insensitive to $\eta_B$, whereas it strongly affects the (unabsorbed) IC high-energy cutoff. Therefore, for a higher $\eta_B$ (stronger $B$) the synchrotron and IC cutoffs get closer. It is worth noting as well that, for a matter dominated jet, it is not expected that $\eta_B$ approaches 1. However, depending on the location of $z_{\rm rec}$ (and thus on the value of $\gamma_{\rm j}$), even values of $\eta_B$ well below one could still mean rather high values for $\eta_B^\star$, yielding an IC flux comparable or even higher than the synchrotron one. As an example, a magnetic fraction $\eta_B = 10^{-2}$ implies $\eta_B^\star \approx 2$ for $\gamma_{\rm j} = 3$, and $\eta_B^\star \approx 0.19$ for $\gamma_{\rm j} = 1.2$. This also means that, regardless of the value of $\eta_{\rm B}$, the strong impact of $z_{\rm rec}$ on the value of $\eta_B^\star$, shows that the synchrotron and the IC predictions are rather sensitive to the parameters determining the wind-jet interaction, which are not well constrained \citep[e.g.,][]{yoon16,bosch16}.

The wind-jet interaction makes our results valid only for distances up to a few $a$. The development of instabilities and significant mixing of the jet and wind material are expected to happen already on the binary system scales \citep{perucho10}. The former could have an important effect on the jet trajectory at larger scales, while the latter would imply a reduction of $\gamma_{\rm j}$ the more the wind material is mixed with the jets. The assumption of a constant $\theta_{\rm j}$-value is also a simplification, as it is likely that the aperture of the jets is modified by the action of the stellar wind \citep{bosch16}. Despite all this, these effects should not have a significant impact on our results as in our model the radiation output tends to be concentrated on a region at $z \lesssim a$ (except for $z_{\rm rec}\sim a$). In any case, proper 3D relativistic simulations of HMMQ jets on large scales would be needed for a more accurate study of the wind-jet interaction.

Both jet instability growth and wind-jet mixing are expected to become more important for clumpy winds \citep{perucho12}, which are usual in massive stars \citep{owocki06,moffat08}. Non-homogeneous winds may also change the accretion rate of material onto the CO by a few 10\% within timescales of several minutes, with a time-averaged value similar to that of an isotropic wind \citep{elmellah18}. This would consequently modify the jet power, introducing also a time variability in $L_{\rm j}$. Moreover, a clumpy wind would also affect the intensity of FFA, given the latter's dependence on the wind density. Considering a simple microclumping model, typical clumping factors $\le 100$ \citep[e.g.,][]{mokiem07,dekoter08} would increase the free-free opacity by a factor of $f \lesssim 10$ \citep{muijres11}, and the wind would remain opaque up to frequencies $f^{1/2} \lesssim 3$ times larger than for a homogeneous wind (see Eq.~\eqref{eq:alphaff}). Therefore, the qualitative results presented in this work regarding the absorption at radio frequencies are independent of the detailed considerations on wind clumping.

Finally, we emphasize that detailed information (e.g., well-sampled light curves) in different energy bands is required in order to significantly constrain the parameter space of the system physical properties. This is the only possible way to disentangle the degeneracy arising from multiple parameter combinations that yield similar results for quantities such as the average flux at a specific energy band. Such studies could be addressed in a future with current and in development observatories, such as \textit{NuSTAR} (hard X-rays), \textit{e-ASTROGAM} (MeV gamma-rays), \textit{Fermi} (high-energy gamma rays), and CTA (very high-energy gamma rays). Devoted observing time with these instruments would help to improve our knowledge on the interplay between the stellar wind and the jets in HMMQs, and the associated non-thermal processes.

\begin{acknowledgements}
We want to thank the referee, Ileyk El Mellah, for his extensive and constructive comments that helped to improve this work. EM and VB acknowledge support by the Spanish Ministerio de Econom\'{i}a y Competitividad (MINECO/FEDER, UE) under grant AYA2016-76012-C3-1-P, with partial support by the European Regional Development Fund (ERDF/FEDER) and MDM-2014-0369 of ICCUB (Unidad de Excelencia `Mar\'{i}a de Maeztu'). EM acknowledges support from MINECO through grant BES-2016-076342. SdP acknowledges support from CONICET. VB acknowledges support from the Catalan DEC grant 2017 SGR 643.
\end{acknowledgements}

\FloatBarrier
\bibliographystyle{aa}
\bibliography{references}

\begin{thebibliography}{52}
\expandafter\ifx\csname natexlab\endcsname\relax\def\natexlab#1{#1}\fi

\bibitem[{{Aharonian}(2004)}]{aharonian04}
{Aharonian}, F.~A. 2004, {Very high energy cosmic gamma radiation: a crucial
  window on the extreme Universe} (World Scientific Publishing Co)

\bibitem[{{Araudo} {et~al.}(2009){Araudo}, {Bosch-Ramon}, \&
  {Romero}}]{araudo09}
{Araudo}, A.~T., {Bosch-Ramon}, V., \& {Romero}, G.~E. 2009, \aap, 503, 673

\bibitem[{{Begelman} {et~al.}(2006){Begelman}, {King}, \&
  {Pringle}}]{begelman06}
{Begelman}, M.~C., {King}, A.~R., \& {Pringle}, J.~E. 2006, \mnras, 370, 399

\bibitem[{{Bosch-Ramon} \& {Barkov}(2016)}]{bosch16}
{Bosch-Ramon}, V. \& {Barkov}, M.~V. 2016, \aap, 590, A119

\bibitem[{{Bosch-Ramon} \& {Khangulyan}(2009)}]{bosch09a}
{Bosch-Ramon}, V. \& {Khangulyan}, D. 2009, Int. J. Mod. Phys. D, 18, 347

\bibitem[{{Crowther} {et~al.}(2006){Crowther}, {Lennon}, \&
  {Walborn}}]{crowther06}
{Crowther}, P.~A., {Lennon}, D.~J., \& {Walborn}, N.~R. 2006, \aap, 446, 279

\bibitem[{{de Koter} {et~al.}(2008){de Koter}, {Vink}, \&
  {Muijres}}]{dekoter08}
{de Koter}, A., {Vink}, J.~S., \& {Muijres}, L. 2008, in Clumping in Hot-Star
  Winds, ed. W.-R. {Hamann}, A.~{Feldmeier}, \& L.~M. {Oskinova}, 47

\bibitem[{{Dermer} \& {Schlickeiser}(2002)}]{dermer02}
{Dermer}, C.~D. \& {Schlickeiser}, R. 2002, \apj, 575, 667

\bibitem[{{Drury}(1983)}]{drury83}
{Drury}, L.~O. 1983, Reports on Progress in Physics, 46, 973

\bibitem[{{Dubus} {et~al.}(2010){Dubus}, {Cerutti}, \& {Henri}}]{dubus10}
{Dubus}, G., {Cerutti}, B., \& {Henri}, G. 2010, \mnras, 404, L55

\bibitem[{{El Mellah} {et~al.}(2019{\natexlab{a}}){El Mellah}, {Sander},
  {Sundqvist}, \& {Keppens}}]{elmellah19a}
{El Mellah}, I., {Sander}, A.~A.~C., {Sundqvist}, J.~O., \& {Keppens}, R.
  2019{\natexlab{a}}, \aap, 622, A189

\bibitem[{{El Mellah} {et~al.}(2018){El Mellah}, {Sundqvist}, \&
  {Keppens}}]{elmellah18}
{El Mellah}, I., {Sundqvist}, J.~O., \& {Keppens}, R. 2018, \mnras, 475, 3240

\bibitem[{{El Mellah} {et~al.}(2019{\natexlab{b}}){El Mellah}, {Sundqvist}, \&
  {Keppens}}]{elmellah19b}
{El Mellah}, I., {Sundqvist}, J.~O., \& {Keppens}, R. 2019{\natexlab{b}}, \aap,
  622, L3

\bibitem[{{Fender} {et~al.}(2004){Fender}, {Belloni}, \& {Gallo}}]{fender04}
{Fender}, R.~P., {Belloni}, T.~M., \& {Gallo}, E. 2004, \mnras, 355, 1105

\bibitem[{{Friend} \& {Castor}(1982)}]{friend82}
{Friend}, D.~B. \& {Castor}, J.~I. 1982, \apj, 261, 293

\bibitem[{{Gies} \& {Bolton}(1986)}]{gies86}
{Gies}, D.~R. \& {Bolton}, C.~T. 1986, \apj, 304, 389

\bibitem[{{Gould} \& {Schr{\'e}der}(1967)}]{gould67}
{Gould}, R.~J. \& {Schr{\'e}der}, G.~P. 1967, Physical Review, 155, 1408

\bibitem[{{Heap} {et~al.}(2006){Heap}, {Lanz}, \& {Hubeny}}]{heap06}
{Heap}, S.~R., {Lanz}, T., \& {Hubeny}, I. 2006, \apj, 638, 409

\bibitem[{{Khangulyan} {et~al.}(2008){Khangulyan}, {Aharonian}, \&
  {Bosch-Ramon}}]{khangulyan08}
{Khangulyan}, D., {Aharonian}, F., \& {Bosch-Ramon}, V. 2008, \mnras, 383, 467

\bibitem[{{Khangulyan} {et~al.}(2014){Khangulyan}, {Aharonian}, \&
  {Kelner}}]{khangulyan14}
{Khangulyan}, D., {Aharonian}, F.~A., \& {Kelner}, S.~R. 2014, \apj, 783, 100

\bibitem[{{Khangulyan} {et~al.}(2018){Khangulyan}, {Bosch-Ramon}, \&
  {Uchiyama}}]{khangulyan18}
{Khangulyan}, D., {Bosch-Ramon}, V., \& {Uchiyama}, Y. 2018, \mnras, 481, 1455

\bibitem[{{Lamers} \& {Cassinelli}(1999)}]{lamers99}
{Lamers}, H.~J.~G.~L.~M. \& {Cassinelli}, J.~P. 1999, {Introduction to Stellar
  Winds} (Cambridge University Press)

\bibitem[{{Leitherer} \& {Robert}(1991)}]{leitherer91}
{Leitherer}, C. \& {Robert}, C. 1991, \apj, 377, 629

\bibitem[{{Longair}(1981)}]{longair81}
{Longair}, M.~S. 1981, {High energy astrophysics} (Cambridge University Press)

\bibitem[{{Luque-Escamilla} {et~al.}(2015){Luque-Escamilla}, {Mart{\'{\i}}}, \&
  {Mart{\'{\i}}nez-Aroza}}]{luque15}
{Luque-Escamilla}, P.~L., {Mart{\'{\i}}}, J., \& {Mart{\'{\i}}nez-Aroza}, J.
  2015, \aap, 584, A122

\bibitem[{{Miller-Jones} {et~al.}(2004){Miller-Jones}, {Blundell}, {Rupen},
  {Mioduszewski}, {Duffy}, \& {Beasley}}]{miller04}
{Miller-Jones}, J.~C.~A., {Blundell}, K.~M., {Rupen}, M.~P., {et~al.} 2004,
  \apj, 600, 368

\bibitem[{{Mioduszewski} {et~al.}(2001){Mioduszewski}, {Rupen}, {Hjellming},
  {Pooley}, \& {Waltman}}]{mioduszewski01}
{Mioduszewski}, A.~J., {Rupen}, M.~P., {Hjellming}, R.~M., {Pooley}, G.~G., \&
  {Waltman}, E.~B. 2001, \apj, 553, 766

\bibitem[{{Moffat}(2008)}]{moffat08}
{Moffat}, A. F.~J. 2008, in Clumping in Hot-Star Winds, ed. W.-R. {Hamann},
  A.~{Feldmeier}, \& L.~M. {Oskinova}, 17

\bibitem[{{Mokiem} {et~al.}(2007){Mokiem}, {de Koter}, {Vink}, {Puls}, {Evans},
  {Smartt}, {Crowther}, {Herrero}, {Langer}, {Lennon}, {Najarro}, \&
  {Villamariz}}]{mokiem07}
{Mokiem}, M.~R., {de Koter}, A., {Vink}, J.~S., {et~al.} 2007, \aap, 473, 603

\bibitem[{{Molina} \& {Bosch-Ramon}(2018)}]{molina18}
{Molina}, E. \& {Bosch-Ramon}, V. 2018, \aap, 618, A146

\bibitem[{{Monceau-Baroux} {et~al.}(2014){Monceau-Baroux}, {Porth}, {Meliani},
  \& {Keppens}}]{monceau2014}
{Monceau-Baroux}, R., {Porth}, O., {Meliani}, Z., \& {Keppens}, R. 2014, \aap,
  561, A30

\bibitem[{{Muijres} {et~al.}(2011){Muijres}, {de Koter}, {Vink},
  {Krti{\v{c}}ka}, {Kub{\'a}t}, \& {Langer}}]{muijres11}
{Muijres}, L.~E., {de Koter}, A., {Vink}, J.~S., {et~al.} 2011, \aap, 526, A32

\bibitem[{{Muijres} {et~al.}(2012){Muijres}, {Vink}, {de Koter}, {M{\"u}ller},
  \& {Langer}}]{muijres12}
{Muijres}, L.~E., {Vink}, J.~S., {de Koter}, A., {M{\"u}ller}, P.~E., \&
  {Langer}, N. 2012, \aap, 537, A37

\bibitem[{{Owocki} \& {Cohen}(2006)}]{owocki06}
{Owocki}, S.~P. \& {Cohen}, D.~H. 2006, \apj, 648, 565

\bibitem[{{Owocki} {et~al.}(2009){Owocki}, {Romero}, {Townsend}, \&
  {Araudo}}]{owocki09}
{Owocki}, S.~P., {Romero}, G.~E., {Townsend}, R.~H.~D., \& {Araudo}, A.~T.
  2009, \apj, 696, 690

\bibitem[{{Pacholczyk}(1970)}]{pacholczyk70}
{Pacholczyk}, A.~G. 1970, {Radio astrophysics. Nonthermal processes in galactic
  and extragalactic sources} (W. H. Freeman \& Company)

\bibitem[{{Pauldrach} {et~al.}(1986){Pauldrach}, {Puls}, \&
  {Kudritzki}}]{pauldrach86}
{Pauldrach}, A., {Puls}, J., \& {Kudritzki}, R.~P. 1986, \aap, 164, 86

\bibitem[{{Perucho} \& {Bosch-Ramon}(2008)}]{perucho08}
{Perucho}, M. \& {Bosch-Ramon}, V. 2008, \aap, 482, 917

\bibitem[{{Perucho} \& {Bosch-Ramon}(2012)}]{perucho12}
{Perucho}, M. \& {Bosch-Ramon}, V. 2012, \aap, 539, A57

\bibitem[{{Perucho} {et~al.}(2010){Perucho}, {Bosch-Ramon}, \&
  {Khangulyan}}]{perucho10}
{Perucho}, M., {Bosch-Ramon}, V., \& {Khangulyan}, D. 2010, \aap, 512, L4

\bibitem[{{Pudritz} {et~al.}(2012){Pudritz}, {Hardcastle}, \&
  {Gabuzda}}]{pudritz12}
{Pudritz}, R.~E., {Hardcastle}, M.~J., \& {Gabuzda}, D.~C. 2012, \ssr, 169, 27

\bibitem[{{Puls} {et~al.}(1996){Puls}, {Kudritzki}, {Herrero}, {Pauldrach},
  {Haser}, {Lennon}, {Gabler}, {Voels}, {Vilchez}, {Wachter}, \&
  {Feldmeier}}]{puls96}
{Puls}, J., {Kudritzki}, R.~P., {Herrero}, A., {et~al.} 1996, \aap, 305, 171

\bibitem[{{Reitberger} {et~al.}(2014){Reitberger}, {Kissmann}, {Reimer}, \&
  {Reimer}}]{reitberger14}
{Reitberger}, K., {Kissmann}, R., {Reimer}, A., \& {Reimer}, O. 2014, \apj,
  789, 87

\bibitem[{{Romero} \& {Orellana}(2005)}]{romero05}
{Romero}, G.~E. \& {Orellana}, M. 2005, \aap, 439, 237

\bibitem[{{Rybicki} \& {Lightman}(1986)}]{rybicki86}
{Rybicki}, G.~B. \& {Lightman}, A.~P. 1986, {Radiative Processes in
  Astrophysics} (Wiley-VCH)

\bibitem[{{Sander} {et~al.}(2018){Sander}, {F{\"u}rst}, {Kretschmar},
  {Oskinova}, {Todt}, {Hainich}, {Shenar}, \& {Hamann}}]{sander18}
{Sander}, A.~A.~C., {F{\"u}rst}, F., {Kretschmar}, P., {et~al.} 2018, \aap,
  610, A60

\bibitem[{{Sikora} {et~al.}(1997){Sikora}, {Madejski}, {Moderski}, \&
  {Poutanen}}]{sikora97}
{Sikora}, M., {Madejski}, G., {Moderski}, R., \& {Poutanen}, J. 1997, \apj,
  484, 108

\bibitem[{{Stirling} {et~al.}(2001){Stirling}, {Spencer}, {de la Force},
  {Garrett}, {Fender}, \& {Ogley}}]{stirling01}
{Stirling}, A.~M., {Spencer}, R.~E., {de la Force}, C.~J., {et~al.} 2001,
  \mnras, 327, 1273

\bibitem[{{Yoon} {et~al.}(2016){Yoon}, {Zdziarski}, \& {Heinz}}]{yoon16}
{Yoon}, D., {Zdziarski}, A.~A., \& {Heinz}, S. 2016, \mnras, 456, 3638

\bibitem[{{Zanin} {et~al.}(2016){Zanin}, {Fern{\'a}ndez-Barral}, {de O{\~n}a
  Wilhelmi}, {Aharonian}, {Blanch}, {Bosch-Ramon}, \& {Galindo}}]{zanin16}
{Zanin}, R., {Fern{\'a}ndez-Barral}, A., {de O{\~n}a Wilhelmi}, E., {et~al.}
  2016, \aap, 596, A55

\bibitem[{{Zdziarski} {et~al.}(2017){Zdziarski}, {Malyshev}, {Chernyakova}, \&
  {Pooley}}]{zdziarski17}
{Zdziarski}, A.~A., {Malyshev}, D., {Chernyakova}, M., \& {Pooley}, G.~G. 2017,
  \mnras, 471, 3657

\bibitem[{{Zdziarski} {et~al.}(2018){Zdziarski}, {Malyshev}, {Dubus}, {Pooley},
  {Johnson}, {Frankowski}, {De Marco}, {Chernyakova}, \& {Rao}}]{zdziarski18}
{Zdziarski}, A.~A., {Malyshev}, D., {Dubus}, G., {et~al.} 2018, \mnras, 479,
  4399

\end{thebibliography}

\end{document}